\documentclass[Journal,10pt]{IEEEtran}
\newcommand{\figref}{Fig.~\ref}

\newcommand{\secref}{Sec.~\ref}

\IEEEoverridecommandlockouts
\usepackage{color}
\usepackage{float}
\usepackage{tabularx,colortbl}
\usepackage{setspace}
\usepackage{cite,graphicx,amsmath,psfrag,bm,amssymb}
\usepackage{subfigure}
\usepackage[usenames,dvipsnames]{xcolor}
\usepackage{mathabx}
\usepackage{epstopdf}
\usepackage{pstool}
\usepackage{caption}
\usepackage{mathtools}
\usepackage{tikz}
\usepackage{xparse}
\DeclareMathOperator{\sinc}{sinc}
\DeclareMathOperator{\rect}{rect}

\definecolor{halfgreen}{rgb}{0,0.5,0}
\definecolor{purplered}{rgb}{0.75,0,0}
\begin{document}
% paper title
\title{Real-Time Dispersion Code Multiple Access (DCMA) for High-Speed Wireless Communications}

\author{Lianfeng Zou,~\IEEEmembership{student member,~IEEE,} Shulabh Gupta,~\IEEEmembership{member,~IEEE,}
        Christophe Caloz,~\IEEEmembership{Fellow,~IEEE}
\thanks{Lianfeng Zou and Christophe Caloz are with the Department
of Electrical Engineering and Poly-Grames Research Center, \'{E}cole Polytechnique de Montr\'{e}al, Montr\'{e}al, Qu\'{e}bec, H3T 1J4, Canada. e-mail: lianfengzou@gmail.com.

Shulabh Gupta is with the Department of Electronics, Carleton University, Ottawa, Ontario, Canada.}%
%\thanks{Manuscript received April 19, 2005; revised December 27, 2012.}
}
\markboth{IEEE TRANSACTIONS ON ANTENNAS AND PROPAGATION,~Vol.~, No.~, Month~2016}%
{Shell \MakeLowercase{\textit{et al.}}: Bare Demo of IEEEtran.cls for Journals}

% make the title area
\maketitle
\begin{abstract}
We model, demonstrate and characterize Dispersion Code Multiple Access (DCMA) and hence show the applicability of this purely analog and real-time multiple access scheme to high-speed wireless communications. We first mathematically describe DCMA and show the appropriateness of Chebyshev dispersion coding in this technology. We next provide an experimental proof-of-concept in a $2\times 2$ DCMA system. Finally, we statistically characterize DCMA in terms of bandwidth, dispersive group delay swing, system dimension and signal-to-noise ratio.
\end{abstract}


%\tableofcontents

\section{Introduction}\label{SEC:INTRO}

High-speed wireless communications require large spectral resources, i.e. ultra wide bandwidth (UWB), available only towards the higher range of the microwave spectrum. For instance, 5G communication systems will operate in the millimeter-wave regime~\cite{Jour:2013_IEEEACCESS_TSRAPPAPORT, Jour:2014_JSAC_JGAndrews,Jour:2014_CommMag_FBoccardi}. Unfortunately, currently existing low-frequency and small-bandwidth standards are not directly transposable to millimeter-wave and THz systems~\cite{CONF:2013_VTC_ERICSSON, JOUR:2014_CommMag_AOsseiran}, and it is therefore highly desirable to explore other UWB technologies.

Impulse Processing (IP) is a particularly promising UWB technology for high-speed wireless communications given its frequency scalability and transceiver simplicity with typical carrier-less transmitter and non-coherent receiver. Combining IP-UWB with an efficient \emph{multiple access technique} allows accommodating more simultaneous users and hence higher spectral efficiency. Conventional IP multiple access techniques include time hopping and frequency hopping. IP time hopping multiple access divides time frames into multiple user chips and distributes these chips according to a time hopping code to minimize error probabilities~\cite{JOUR:1998_CommLett_Win, JOUR:2000_TCOM_Win}. However, this technique typically requires digital processing, which impairs its frequency scalability. A purely analog IP frequency hopping multiple access has been exclusively used in optics so far. This analog technique consists in rearranging the spectral contents of signals in time, using dispersion controlled devices such as Bragg fiber gratings~\cite{JOUR:1997_JLT_Hill}, so as to encode ultra-short pulses according to a dispersive time-frequency mapping scheme~\cite{JOUR:1999_JLT_Fathallah, JOUR:2001_PTL_Chen, JOUR:2004_PTL_Tamai}. The purely analog nature of this technique attractively leads to real-time processing, and hence low signal latency and high-speed transmission. However, its application to radio signal typically requires a microwave photonics approach~\cite{JOUR:2009_JLT_JPYAO}, which involves lossy, complex, expensive and non-integrable electro-optical technology.

Radio Real-time Analog Signal Processing (RR-ASP) represents a promising alternative to IP-UWB technology for wireless communications~\cite{Jour:2013_MwMag_Caloz}. The core of an RR-ASP is the ``phaser", which is a dispersion engineered device that provides specified group delay versus frequency response, $\tau(\omega)$. A phaser transforms a UWB pulse into a time-spread version of it with spectral components rearranged in time. Practical phasers may be realized in various technologies, such as all-pass~\cite{JOUR:1966_TMTT_Cristal,JOUR:2010_TMTT_Gupta,JOUR:2015_TMTT_Gupta}, coupled-resonator~\cite{JOUR:2012_TMTT_Zhang, JOUR:2013_TMTT_QZhang}, tapped delay line~\cite{JOUR:2013_TMTT_Xiang,Jour:2015_TAP_Ding}, and electromagnetic band gap (EBG)~\cite{JOUR:2003_TMTT_Laso} technologies. Many RR-ASP applications have already been reported, including real-time Fourier transformation~\cite{JOUR:2003_TMTT_Laso}, frequency sniffing~\cite{JOUR:2012_MWCL_Nikfal}, pulse compression, expansion and time-reversal~\cite{JOUR:2007_TMTT_Schwartz, CONF:2008_IRWS_Schwartz}, frequency division multiplexing~\cite{CONF:2011_EuMC_Nikfal}, signal-to-noise ratio enhancement~\cite{JOUR:2014_MWCL_Nikfal}, and uniform radiation scanning for antenna array~\cite{CONF:2015_APS_Zhang}, to name a few.

Dispersion Code Multiple Access (DCMA) has been reported in~\cite{CONF:2015_APS_Gupta,CONF:2016_APS_ZOU} as an efficient RR-ASP UWB multiple access technique. DCMA may be considered as the wireless and radio counterpart of Bragg grating based multiple access~\cite{JOUR:1999_JLT_Fathallah} in optics. This paper presents an in-depth characterization of DCMA in an random line-of-sight (LOS) wireless environment and demonstrates an experimental proof-of-concept of a DCMA system. Only LOS propagation, essentially corresponding to an open or uncluttered environment, is characterized. Non-LOS characterization, better describing a close and cluttered environment, will be reported elsewhere.

The paper is organized as follows. First, \secref{SEC:DCMA} conceptually and mathematically describes a general DCMA system. Next, \secref{SEC:DCSEL} introduces the Chebyshev phasing as a convenient coding scheme for DCMA. Then, \secref{SEC:EXPERIMENT} provides the experimental proof-of-concept of a DCMA system in a LOS configuration. From this point, \secref{SEC:SYS} characterizes a general DCMA system in an arbitrary LOS channel. Conclusions are given in \secref{SEC:CONCL}.

\section{DCMA Concept}\label{SEC:DCMA}

\subsection{Principle of Operation and Notation}\label{SEC:DCMA:CONCEPT}

\begin{figure}[h!t]
   \centering
   \includegraphics[width=1\columnwidth]{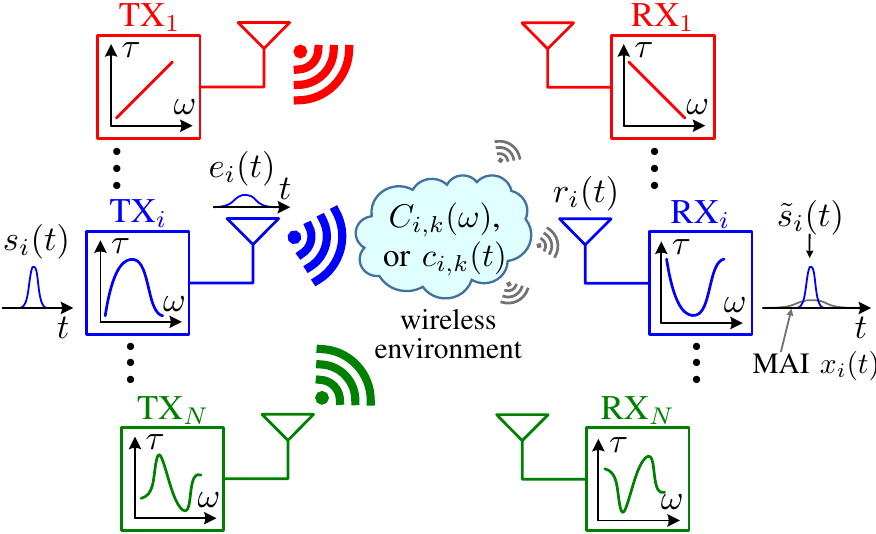}
%   \psfragfig[width=1\linewidth, trim={0in 0in 0in 0in}]{dcmconcept}{
%        \psfrag{t}[r][r][1]{$t$}
%        \psfrag{W}[c][c][0.85]{wireless}
%        \psfrag{C}[cb][cm][0.85]{environment}
%        \psfrag{H}[c][c][0.95]{$C_{ik}(\omega)$,}
%        \psfrag{G}[c][c][0.95]{or $c_{ik}(t) $}
%       \psfrag{g}[l][l][1]{$\tau$}
%       \psfrag{w}[r][r][1]{$\omega$}
%       \psfrag{I}[l][l][0.85]{MAI $x_i(t)$}
%       \psfrag{s}[l][l][1]{$s_i(t)$}
%       \psfrag{e}[l][l][1]{$e_i(t)$}
%       \psfrag{r}[l][l][1]{$r_i(t)$}
%       \psfrag{z}[c][c][1]{$\tilde{s}_i(t)$}
%       \psfrag{1}[c][c][1]{\textcolor{red}{TX$_1$}}
%       \psfrag{2}[c][c][1]{\textcolor{blue}{TX$_i$}}
%       \psfrag{3}[c][c][1]{\textcolor{halfgreen}{TX$_N$}}
%       \psfrag{4}[c][c][1]{\textcolor{red}{RX$_1$}}
%       \psfrag{5}[c][c][1]{\textcolor{blue}{RX$_i$}}
%       \psfrag{6}[c][c][1]{\textcolor{halfgreen}{RX$_N$}}
%   }
   \caption{Schematic representation of an $N\times N$ dispersion code multiple access (DCMA) communication system.}
   \label{FIG:DCMA_CONCEPT}
\end{figure}

Figure~\ref{FIG:DCMA_CONCEPT} shows the conceptual schematic of a DCMA system composed of $N$ transmitter-receiver (TX-RX) pairs, which may be placed anywhere. The signal sent from TX$_i$ is a modulated UWB pulse train, $s_i(t)$. This signal is encoded by the corresponding phaser into the new signal $e_i(t)$, which is intended to be received by RX$_i$ as $r_i(t)$. To ensure such transmission from TX$_i$ to RX$_i$, the corresponding encoding and decoding phaser transfer functions must be assigned a pair of group delay frequency functions (dispersion codes), $\tau_{\text{TX}i}(\omega)$ and $\tau_{\text{RX}i}(\omega)$, that are matched (or phase-conjugated), i.e.
\begin{equation}\label{EQ:DCMCONCEPT_MATCHED_CONDITION}
  \tau_{\text{TX}i}(\omega)+\tau_{\text{RX}i}(\omega) =  \text{constant}.
\end{equation}

The system operates in a wireless environment, whose channel, that is generally based on statistical model, may be described by the transfer function
\begin{equation}\label{EQ:DCMCONCEPT_CHAN_TRANSFUNC}
 C_{ik}(\omega) = \mathcal{F}[c_{ik}(t)],\quad i,k\in\{1,2,\ldots,N\},
\end{equation}
between TX$_k$ and RX$_i$, where $c_{ik}(t)$ is the corresponding impulse response and $\mathcal{F}$ represents the Fourier transform operation.

When all of the $N$ pairs are simultaneously active, the decoding phaser at RX$_i$ recovers the desired signal, $\tilde{s}_i(t)$, which is a distorted replica of $s_i(t)$ due to both channel fading and interference originating from the $N-1$ undesired transmitters, $x_i(t)$, called multiple access interference (MAI).

\subsection{Mathematical Description}\label{SEC:DCMA:MATH}

In this section, we shall provide the mathematical description of the overall DCMA system operation. A matched pair of dispersion codes, corresponding to~\eqref{EQ:DCMCONCEPT_MATCHED_CONDITION}, may be expressed as
\begin{subequations}\label{EQ:DCMCONCEPT_MATCHED_GD}
\begin{equation}\label{EQ:DCMCONCEPT_MATCHED_GDTX}
    \tau_{\text{TX}_i}(\omega)  = \tau_0 + \tau_i(\omega),
\end{equation}
\begin{equation}\label{EQ:DCMCONCEPT_MATCHED_GDRX}
    \tau_{\text{RX}_i}(\omega)  = \tau_0 - \tau_i(\omega).
\end{equation}
\end{subequations}
Assuming bandwidth-limited (e.g. bandpass) and lossless phasers, the corresponding transfer functions may be written
\begin{subequations}\label{EQ:DCMCONCEPT_MATCHED_TF}
  \begin{equation}\label{EQ:DCMCONCEPT_MATCHED_TFTX}
    H_{\text{TX}_i}(\omega) = \rect\left(\dfrac{\omega-\omega_0}{\Delta\omega}\right)e^{-j\omega\tau_0}e^{j\phi_i(\omega)},
  \end{equation}
  \begin{equation}\label{EQ:DCMCONCEPT_MATCHED_TFRX}
    H_{\text{RX}_i}(\omega) = \rect\left(\dfrac{\omega-\omega_0}{\Delta\omega}\right)e^{-j\omega\tau_0}e^{-j\phi_i(\omega)},
  \end{equation}
\end{subequations}
where $\phi_i(\omega)=-\int_{\omega_0-\Delta\omega/2}^{\omega}\tau_i(\omega')\,d\omega'$ is a dispersive (non-linear) phase function of frequency, and $\omega_0=2\pi f_0$ and $\Delta\omega = 2\pi \Delta f$ are center frequency and bandwidth, respectively, of the system. %The corresponding impulse responses are

The signal arriving at RX$_i$ is
\begin{equation}\label{EQ:DCMCONCEPT_RECSIG}
  R_i(\omega) = \left(\sum_{k=1}^{N} E_kC_{ik}\right)(\omega)
             =  \left(\sum_{k=1}^{N}S_kH_{\text{TX}_k}C_{ik}\right)(\omega)+N(\omega),
\end{equation}
which is the sum of the encoded signals from all the transmitters propagated through their respective channels plus noise $N(\omega)$. One finds the decoded signal by multiplying~\eqref{EQ:DCMCONCEPT_RECSIG} with $H_{\text{RX}_i}(\omega)$, i.e.
\begin{subequations}
\begin{equation}\label{EQ:DCMCONCEPT_DECSIGSPC}
\begin{split}
   Z_i(\omega) &= H_{\text{RX}_i}(\omega)\left(\sum_{k=1}^{N}S_kH_{\text{TX}_k}C_{ik}\right)(\omega)+N(\omega)\\
   & = \tilde{S}_i(\omega) + X_i(\omega) + N(\omega),
\end{split}
\end{equation}
\noindent with
\begin{equation}\label{EQ:DCMCONCEPT_DECSIGSPC_DESIRED}
   \tilde{S}_i(\omega) = \left(S_iH_{ii}C_{ii}\right)(\omega),
\end{equation}
\begin{equation}\label{EQ:DCMCONCEPT_DECSIGSPC_MAI}
  X_i(\omega) = \left[\sum_{\substack{k=1\\k\neq i}}^N S_k H_{ik}C_{ik}\right](\omega).
\end{equation}
\end{subequations}
The cascaded transfer functions (excluding the channel response) from TX$_i$ to RX$_i$, $H_{ii}(\omega)$ and from TX$_k$ to RX$_i$, $H_{ik}(\omega)$ are the product of~\eqref{EQ:DCMCONCEPT_MATCHED_TFRX} and~\eqref{EQ:DCMCONCEPT_MATCHED_TFTX} with $i$ replaced by $k$ for $H_{ik}(\omega)$, leading to
\begin{subequations}
\begin{equation}\label{EQ:DCMCONCEPT_CASTFII}
  H_{ii}(\omega) = (H_{\text{TX}_i} H_{\text{RX}_i})(\omega) = \rect\left(\dfrac{\omega-\omega_0}{\Delta\omega}\right)e^{-j\omega(2\tau_0)},
\end{equation}
\begin{equation}\label{EQ:DCMCONCEPT_CASTF}
  H_{ik}(\omega) = (H_{\text{TX}_k} H_{\text{RX}_i})(\omega) =\rect\left(\dfrac{\omega-\omega_0}{\Delta\omega}\right)e^{j\phi_{ik}(\omega)},
\end{equation}
\end{subequations}
with cascaded phase
\begin{equation}\label{EQ:DCMCONCEPT_CASPHS}
  \phi_{ik}(\omega) = -\int_{\omega_0-\Delta\omega/2}^{\omega}\tau_{ik}(\omega')d\omega',\quad k\neq i,
\end{equation}
and the cascaded group delay $\tau_{ik}(\omega)$ is the sum of~\eqref{EQ:DCMCONCEPT_MATCHED_GDRX} and~\eqref{EQ:DCMCONCEPT_MATCHED_GDTX} for $i$ replaced by $k$, i.e.
\begin{equation}\label{EQ:DCMCONCEPT_CASGD}
  \tau_{ik}(\omega) = \tau_{\text{TX}_k}(\omega)+\tau_{\text{RX}_i}(\omega) = 2\tau_0 +\tau_k(\omega)-\tau_i(\omega),~k\neq i.
\end{equation}
 The corresponding cascaded impulse responses are then
\begin{subequations}\label{EQ:DCMCONCEPT_CASIRs}
  \begin{equation}\label{EQ:DCMCONCEPT_AUTOCORR}
  \begin{split}
      h_{ii}(t) &= \mathcal{F}^{-1}\left[ H_{ii}(\omega)\right]\\
      & = 2\Delta f \dfrac{\sin[\pi\Delta f (t-2\tau_0)]}{\pi\Delta f (t-2\tau_0)} \cos[2\pi f_0(t-2\tau_0)],
  \end{split}
\end{equation}
whose envelope is the $\sinc$ function and peak amplitude, at $t=2\tau_0$, is proportional to the system bandwidth ($\Delta f$), and
\begin{equation}\label{EQ:DCMCONCEPT_CASIR}
    h_{ik}(t) =\mathcal{F}^{-1}\left[ H_{ik}(\omega)\right] = (h_{\text{TX}_k}\ast h_{\text{RX}_i})(t),
\end{equation}
\end{subequations}
where ``$\ast$" denotes the convolution product, and $h_{\text{TX}_k}(t)$ $h_{\text{RX}_i}(t)$ are the impulse responses of the corresponding encoding and decoding phasers, respectively.

The waveform corresponding to $Z_i(\omega)$ in~\eqref{EQ:DCMCONCEPT_DECSIGSPC} is thus
\begin{subequations}\label{EQ:DCMCONCEPT_DECSIGWFMTOT}
  \begin{equation}\label{EQ:DCMCONCEPT_DECSIGWFM}
   {z}_i(t) = \mathcal{F}^{-1}\left[Z_i(\omega)\right] = \tilde{s}_i(t)+x_i(t)+ n(t),
\end{equation}
with
\begin{equation}\label{EQ:DCMCONCEPT_DECSIGWFM_DESIRED}
  \tilde{s}_i(t) = \mathcal{F}^{-1}\left[\tilde{S}_i(\omega)\right] = s_i(t)\ast h_{ii}(t)\ast c_{ii}(t),
\end{equation}
\begin{equation}\label{EQ:DCMCONCEPT_DECSIGWFM_MAI}
  x_i(t) = \mathcal{F}^{-1}\left[X_i(\omega)\right] = \sum_{\substack{k=1\\k\neq i}}^N s_k(t)\ast h_{ik}(t)\ast c_{ik}(t),
\end{equation}
and $n(t)$ typically being Additive White Gaussian Noise (AWGN), whose time sample value
\begin{equation}\label{EQ:DCMCONCEPT_DECSIGWFM_AWGN}
  n =\mathcal{N}\left(0, \sigma_\text{N}^2\right),
\end{equation}
\end{subequations}
where $\mathcal{N}(\cdot)$ denotes the normal distribution with zero mean and standard deviation $\sigma_\text{N}$ corresponding to the noise power $\sigma_\text{N}^2$.
Throughout the paper, all the plotted impulse responses and time-domain signals are normalized by the peak of $h_{ii}(t)$, $2\Delta f$, as
\begin{equation}\label{EQ:DCMCONCEPT_NORM}
\begin{split}
    &\hat{h}_{ik}(t) = \dfrac{{h}_{ik}(t)}{2\Delta f},~\forall\,i,k,\quad \hat{z}_i(t) =\dfrac{{z}_i(t)}{2\Delta f},\\
    &\hat{\tilde{s}}_i(t) = \dfrac{{\tilde{s}}_i(t)}{2\Delta f},~ \hat{x}_i(t)=\dfrac{{x}_i(t)}{2\Delta f},~\hat n(t) = \dfrac{n(t)}{2\Delta f}.
\end{split}
\end{equation}

Successful transmission requires that $z_i(t)=\tilde{s}_i(t)+x_i(t)+ n(t)$ be higher than a specified threshold if the transmitted bit is a 1 and $z_i(t)=x_i(t)+ n(t)$ be lower than that threshold if the transmitted bit is a 0.

\section{Chebyshev Dispersion Coding Functions}\label{SEC:DCSEL}

\subsection{Chebyshev Coding Selection}

In addition to phase conjugation in user pairs, expressed in~\eqref{EQ:DCMCONCEPT_MATCHED_CONDITION} and enforced in~\eqref{EQ:DCMCONCEPT_MATCHED_TF}, DCMA requires dispersion code diversity to accommodate multiple user pairs. A set of appropriate functions should therefore be selected as dispersion codes.
In multiple access technologies based on coherent detection schemes, such as for instance OFDMA or CDMA~\cite{BK:2011_MOLISCH_WIRELESSCOMM}, the optimal coding functions form orthogonal sets, and corresponding MAI signals are ideally not affecting detection. However, as pointed out in Sec.~\ref{SEC:INTRO}, DCMA is fundamentally based on non-coherent (threshold) detection. While being an advantage in terms of simplicity and cost, this fact makes orthogonal sets irrelevant in typical wireless channels, and other dispersion coding functions must therefore be identified.

In~\cite{CONF:2015_APS_Gupta}, we used Chebyshev polynomials of the first kind, $T_m(x)$, as the dispersion coding functions, because these polynomials all exhibit identical peak-to-peak amplitudes ($-1$ to $+1$) as $x$ varies between $-1$ and $+1$~\cite{BK:2007_Andrei}, leading to identical maximal MAI [$x_i(t)$] time spreads in all the receivers. These identical time spreads correspond to identical average MAI powers, as will be shown later in this section. Another reason for this choice is the fact that Chebyschev phasers have recently become available, in non-uniform transmission-line C-section technology~\cite{JOUR:2016_MWCL_STARAVATI}. Note that the dispersion codes used in optical multiple access are typically stair-case group delay functions~\cite{JOUR:1999_JLT_Fathallah, JOUR:2001_PTL_Chen, JOUR:2004_PTL_Tamai}. However,  these functions involve broad idle frequency guard bands between adjacent group delay stairs, due to the sinc-type frequency slicing by Bragg grating fibers, which represents a waste of spectral resources that is unaffordable at RF in addition to extra constraints on the phasers design.

\begin{figure}[h!t]
   \centering
     \includegraphics[width=1\columnwidth]{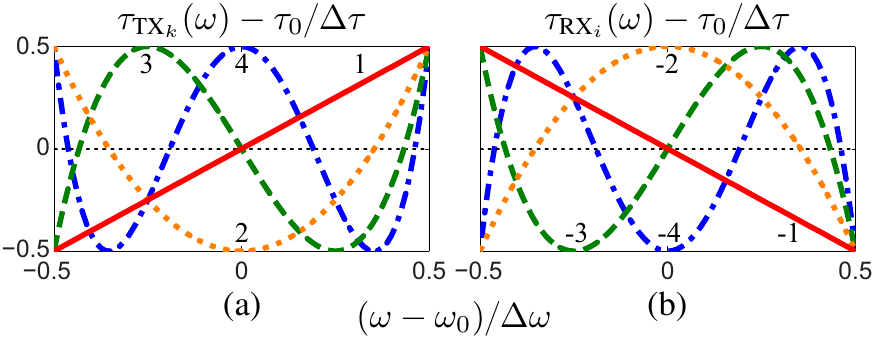}
%   \psfragfig[width=1\linewidth, trim={0in -0.13in 0in -0.15in}]{dcexample}{
%   \psfrag{T}[cb][cb][1]{${\tau_{\text{TX}_k}(\omega)-\tau_0}/{\Delta\tau}$}
%   \psfrag{R}[cb][cb][1]{${\tau_{\text{RX}_i}(\omega)-\tau_0}/{\Delta\tau}$}
%   \psfrag{f}[ct][c][1]{$(\omega-\omega_0)/\Delta\omega$}
%   \psfrag{A}[c][c][1]{(a)}
%   \psfrag{B}[c][c][1]{(b)}
%   }
   \caption{First few normalized Chebyshev dispersion function coding pairs, corresponding to $\mathbf{C}=[1,2,3,4]$ with group delay swing $\Delta\tau$ and bandwidth $\Delta f$. (a)~Encoding codes. (b)~Corresponding (phase-conjugated) decoding codes.}
   \label{FIG:DCSEL_CHEBYGD_4CHGDS}
\end{figure}

For a given group delay swing of $\Delta\tau=\tau_\text{max} - \tau_\text{min}$, the $i^\text{th}$ Chebyschev dispersion coding function pair may by written as
\begin{subequations}\label{EQ:DCSEL_CHEBY_GD}
\begin{equation}\label{EQ:DCSEL_CHEBY_GD_TX}
    \tau_{\text{TX}_i}(\omega) = \tau_0+\dfrac{\Delta\tau}{2} T_{m_i}\left(\dfrac{\omega-\omega_0}{\Delta\omega/2}\right),
\end{equation}
\begin{equation}\label{EQ:DCSEL_CHEBY_GD_RX}
      \tau_{\text{RX}_i}(\omega) = \tau_0-\dfrac{\Delta\tau}{2} T_{m_i}\left(\dfrac{\omega-\omega_0}{\Delta\omega/2}\right),
\end{equation}
\end{subequations}
where
\begin{equation*}
   T_{m_i}(x) =  \cos(m_i\arccos x),\quad m_i\in\mathbb{Z}\;
  \text{and}\; x\in[-1,1],
\end{equation*}
is the $ m_i^\text{th}$ order Chebyshev polynomial of the first kind. Then, one may use the encoding set
\begin{equation}\label{EQ:DCSEL_CHEBY_CODESET}
  \mathbf{C} = [m_1,\ldots, m_i,\ldots,m_N],
  \quad m_i\neq m_k,
\end{equation}
whose elements may take any integer values in any order with the only restriction that all these values appear only once. For convenience, we define $-T_{m_i}(x)=T_{-m_i}(x)$ ($m_i>0$) for the phase-conjugated functions. Figure~\ref{FIG:DCSEL_CHEBYGD_4CHGDS} plots the first four Chebyshev group delay functions, corresponding to $\mathbf{C} = [1,2,3,4]$.

\subsection{Delay -- Waveform Relationships}

In order to better understand the relationships between the group delay function coding pairs~\eqref{EQ:DCSEL_CHEBY_GD} and the related DCMA waveforms, we shall now investigate a few important coding sets. We shall assume, for simplicity, the scenario where \mbox{1)~$s_i(t)=\delta(t-t_{\text{Tx}_i})$}, i.e. the excitation signal energy is flat across the system bandwidth ($\Delta f$), 2)~$t_{\text{Tx}_i}=t_{\text{Tx}_k}=0,\,\forall\,i,k$, i.e. all the transmitters are synchronized, 3)~$c_{ik}(t)=\delta(t),\,\forall\,i,k$, i.e. all the channels are identical and shorted, and 4)~$n(t)=0$, i.e. no noise is present. In this scenario, Eqs.~\eqref{EQ:DCMCONCEPT_DECSIGWFMTOT} reduce to
\begin{equation}\label{EQ:DCSEL_DECSIG}
  \tilde{s}_i(t) = h_{ii}(t),~x_i(t) = \sum_{\substack{k=1\\k\neq i}}^N h_{ik}(t),~z_i(t) = \tilde{s}_i(t) + x_i(t).
\end{equation}

\begin{figure}[h!t]
   \centering
   \subfigure[]{\label{FIG:GD_GENERAL}
   \includegraphics[width=1\columnwidth]{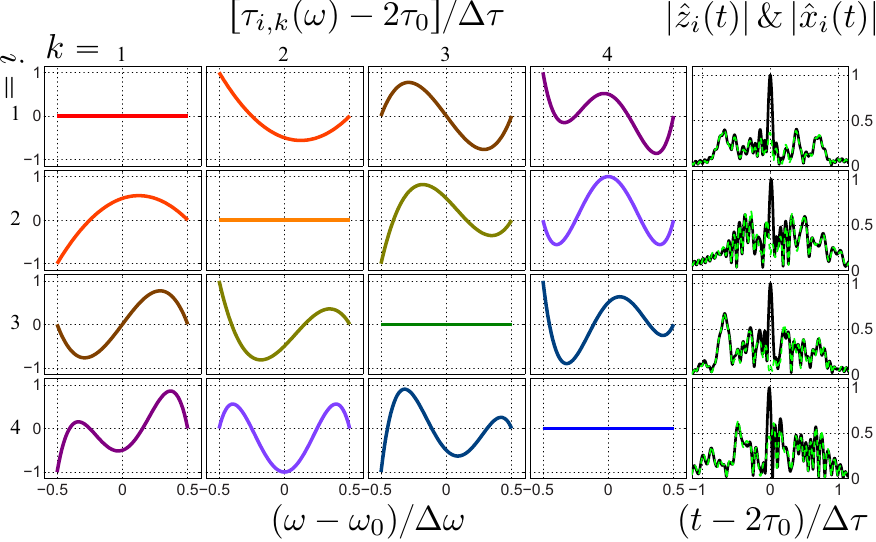}
   %\psfragfig[width=1\linewidth, trim={-0.06in -0.3in 0in -0.25in}]{gdarraygeneral}{
%   \psfrag{K}[bl][bl][1]{$k=$}
%   \psfrag{I}[l][lt][1]{$i=$}
%   \psfrag{F}[ct][ct][1]{$(\omega-\omega_0)/\Delta\omega$}
%   \psfrag{T}[ct][ct][1]{$(t-2\tau_0)/\Delta\tau$}
%   \psfrag{D}[cb][cb][1]{$[\tau_{ik}(\omega)-2\tau_0]/\Delta\tau$}
%   \psfrag{S}[cb][cb][1]{$|\hat{z}_i(t)|\, \&\, |\hat{x}_i(t)|$}
%   }
}
   \subfigure[]{\label{FIG:GD_ODD}
      \includegraphics[width=1\columnwidth]{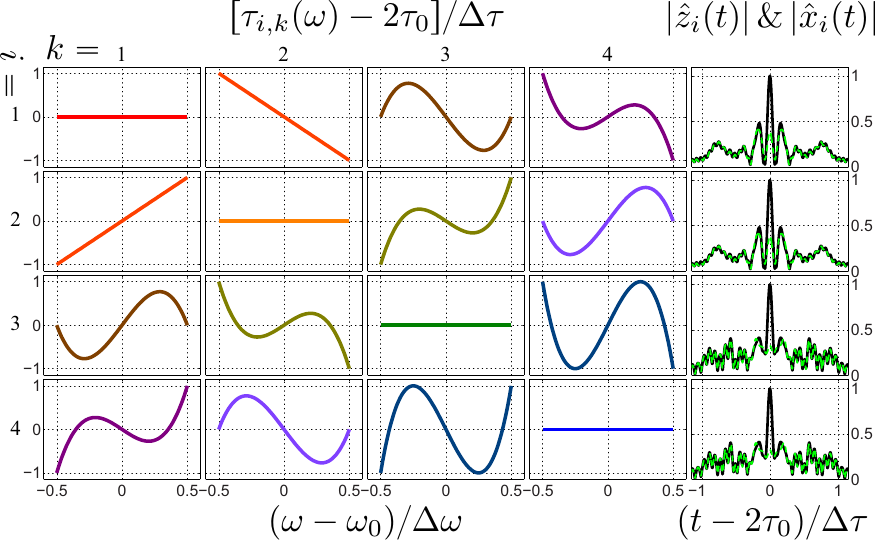}
 %  \psfragfig[width=1\linewidth, trim={-0.06in -0.3in 0in -0.25in}]{gdarrayodd}{
%   \psfrag{K}[bl][bl][1]{$k=$}
%   \psfrag{I}[l][l][1]{$i=$}
%   \psfrag{F}[ct][ct][1]{$(\omega-\omega_0)/\Delta\omega$}
%   \psfrag{T}[ct][ct][1]{$(t-2\tau_0)/\Delta\tau$}
%   \psfrag{D}[cb][cb][1]{$[\tau_{ik}(\omega)-2\tau_0]/\Delta\tau$}
%   \psfrag{S}[cb][cb][1]{$|\hat{z}_i(t)|\, \&\, |\hat{x}_i(t)|$}
%   }
}
   \subfigure[]{\label{FIG:GD_EVEN}
      \includegraphics[width=1\columnwidth]{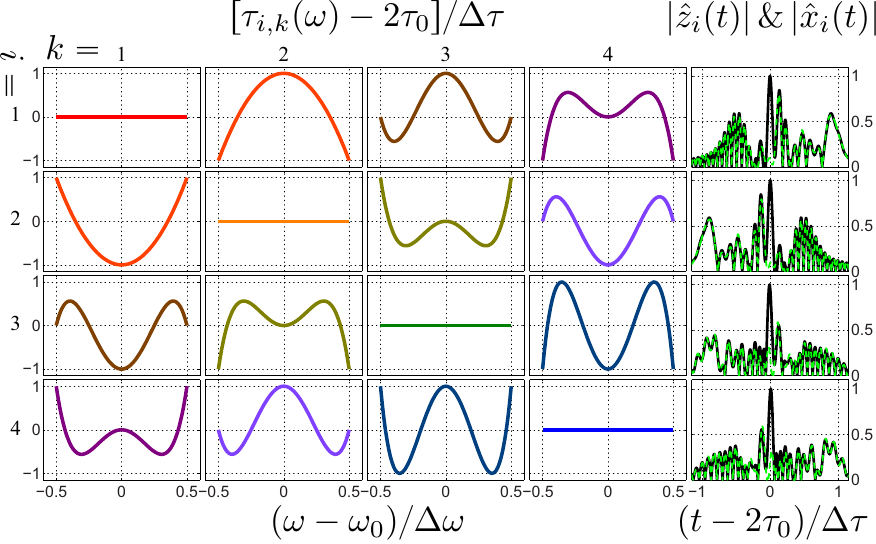}
%   \psfragfig[width=1\linewidth, trim={-0.06in -0.3in 0in -0.25in}]{gdarrayeven}{
%   \psfrag{K}[bl][bl][1]{$k=$}
%   \psfrag{I}[l][l][1]{$i=$}
%   \psfrag{F}[ct][ct][1]{$(\omega-\omega_0)/\Delta\omega$}
%   \psfrag{T}[ct][ct][1]{$(t-2\tau_0)/\Delta\tau$}
%   \psfrag{D}[cb][cb][1]{$[\tau_{ik}(\omega)-2\tau_0]/\Delta\tau$}
%   \psfrag{S}[cb][cb][1]{$|\hat{z}_i(t)|\, \&\, |\hat{x}_i(t)|$}
%   }
}
   \caption{Normalized cascaded group delay functions, $\tau_{ik}(\omega)$ [Eq.~\eqref{EQ:DCMCONCEPT_CASGD} with~\eqref{EQ:DCSEL_CHEBY_GD}] (first four columns) and signal envelopes obtained by the inverse Fourier transform of~\eqref{EQ:DCMCONCEPT_CASIR} with~\eqref{EQ:DCMCONCEPT_CASTF} and~\eqref{EQ:DCSEL_CHEBY_GD}, $\Delta\tau=4$~ns and $\Delta f=4$~GHz] (last column) with decoded signal evelope, $|\hat{z}_i(t)|$, in solid black lines and MAI, $|\hat{x}_i(t)|$, in dashed green lines. (a)~Coding set corresponding to~\figref{FIG:DCSEL_CHEBYGD_4CHGDS} (mixture of even and odd functions). (b)~All-odd coding set, $\mathbf{C} = [1,-1,3,-3]$ (odd functions). (c)~All-even coding set: $\mathbf{C} = [2,-2,4,-4]$ (even functions). It is seen that the MAI peak level is lowest in~(b), intermediate in~(a) and highest in~(c).}
   \label{FIG:CASGD_AND_WFM}
\end{figure}

Figure~\ref{FIG:CASGD_AND_WFM} plots the cascaded group delay pairs $\tau_{ik}(\omega)$ (first four columns), and the decoded waveform envelope $|\hat{z}_i(t)|$ and corresponding MAI envelope $|\hat{x}_i(t)|$ (last column). In the first four columns, the diagonal graphs ($k=i$) are matched cascaded group delays, $\tau_{ii}(\omega)$, which are constant over the entire system bandwidth. These group delays correspond to the desired signals $\tilde{s}_i(t) =h_{ii}(t)$ analytically given by~\eqref{EQ:DCMCONCEPT_AUTOCORR}. The off-diagonal graphs ($k\neq i$) are unmatched cascaded group delays, $\tau_{ik}(\omega)$, corresponding to MAI, $h_{ik}(t)$, from TX$_k$. The corresponding group delay swing is
\begin{equation}\label{EQ:DCSEL_GDSWING_IK}
  \Delta\tau_{ik}=\max\left[\tau_{ik}(\omega)\right]-\min\left[\tau_{ik}(\omega)\right]\leq 2\Delta\tau,
\end{equation}
which is approximately equal to the spread of $h_{ik}(t)$.

Figure~\ref{FIG:GD_GENERAL} corresponds to the coding set in~\figref{FIG:DCSEL_CHEBYGD_4CHGDS}, i.e. $\mathbf{C}=[1,2,3,4]$, including thus mixed odd and even Chebyshev functions, while Figs.~\ref{FIG:GD_ODD} and~\ref{FIG:GD_EVEN} correspond to the cases of all-odd and all-even Chebyshev functions, $\mathbf{C}=[1,-1,3,-3]$ and $\mathbf{C}=[2,-2,4,-4]$, respectively. An overall comparison of the corresponding signal waveforms (last column sets) reveals that the MAI peak level and hence the corresponding bit error probability (BEP), is lowest for the all-odd set [\figref{FIG:GD_ODD}], intermediate for the mixed odd-even set [\figref{FIG:GD_GENERAL}] and highest for the all-even set [\figref{FIG:GD_EVEN}], as first noted in~\cite{CONF:2015_APS_Gupta}. Moreover, one may observe a key difference of group delay profiles between the all-odd [Fig.~\ref{FIG:GD_ODD}] and all-even [Fig.~\ref{FIG:GD_EVEN}] coding sets. The group delay profiles of the former are anti-symmetric about the center time and center frequency, while those of the latter are symmetric about the center frequency. These symmetries are at the origin of the results, as shall be qualitatively explained next.

\begin{figure}[h!t]
   \centering
    \subfigure[]{\label{FIG:GDWFMS_ODD}
    \includegraphics[width=1\columnwidth]{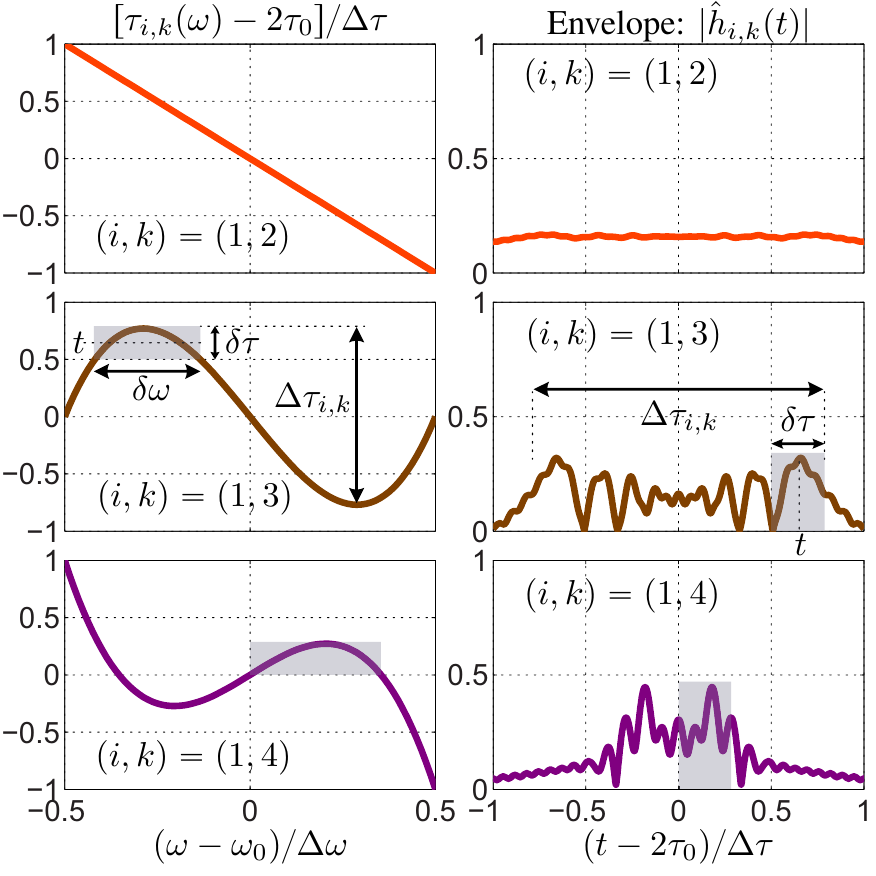}
%   \psfragfig[width=1\linewidth, trim={0in 0in 0in 0in}]{gdwfmsodd}{
%   \psfrag{A}[l][l][1]{$(i,k)=(1,2)$}
%   \psfrag{B}[l][l][1]{$(i,k)=(1,3)$}
%   \psfrag{C}[l][l][1]{$(i,k)=(1,4)$}
%   \psfrag{T}[c][c][1]{$(t-2\tau_0)/\Delta\tau$}
%   \psfrag{F}[c][c][1]{$(\omega-\omega_0)/\Delta\omega$}
%   \psfrag{G}[c][c][1]{$[\tau_{ik}(\omega)-2\tau_0]/\Delta\tau$}
%   \psfrag{W}[c][c][1]{Envelope: $|\hat{h}_{ik}(t)|$}
%   \psfrag{D}[c][c][1]{$\delta\omega$}
%   \psfrag{E}[l][l][1]{$\delta\tau$}
%   \psfrag{H}[r][r][1]{$\Delta\tau_{ik}$}
%   \psfrag{a}[r][r][1]{$t$}
%   }
}
    \subfigure[]{\label{FIG:GDWFMS_EVEN}
    \includegraphics[width=1\columnwidth]{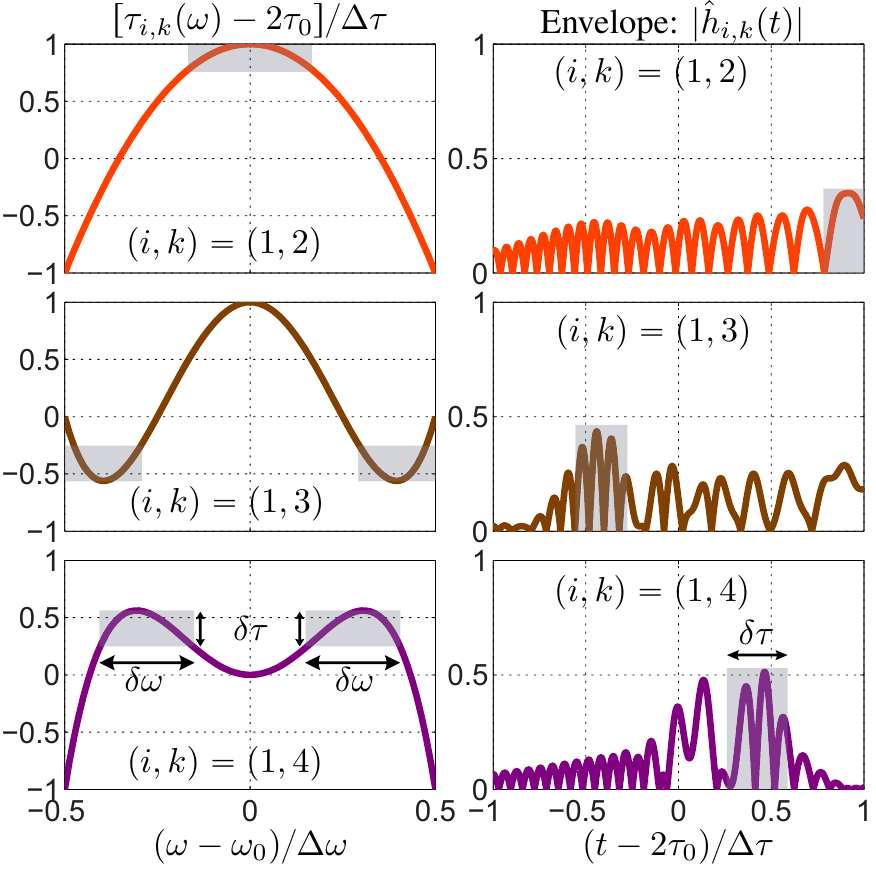}
%   \psfragfig[width=1\linewidth, trim={-0.00in 0in 0in 0in}]{gdwfmseven}{
%   \psfrag{A}[l][l][1]{$(i,k)=(1,2)$}
%   \psfrag{B}[l][l][1]{$(i,k)=(1,3)$}
%   \psfrag{C}[l][l][1]{$(i,k)=(1,4)$}
%   \psfrag{T}[c][c][1]{$(t-2\tau_0)/\Delta\tau$}
%   \psfrag{F}[c][c][1]{$(\omega-\omega_0)/\Delta\omega$}
%   \psfrag{G}[c][c][1]{$[\tau_{ik}(\omega)-2\tau_0]/\Delta\tau$}
%   \psfrag{W}[c][c][1]{Envelope: $|\hat{h}_{ik}(t)|$ }
%   \psfrag{D}[c][c][1]{$\delta\omega$}
%   \psfrag{E}[c][c][1]{$\delta\tau$}
%   }
}
   \caption{Graphical explanation of the relationships between the cascaded group delay sets $\tau_{ik}(\omega)$ [Eq.~\eqref{EQ:DCMCONCEPT_CASGD} with~\eqref{EQ:DCSEL_CHEBY_GD}] in~\figref{FIG:CASGD_AND_WFM} and MAI contribution, $|\hat{h}_{ik}(t)|$ (envelope) [inverse Fourier transform of Eq.~\eqref{EQ:DCMCONCEPT_CASIR} with~\eqref{EQ:DCMCONCEPT_CASTF} and~\eqref{EQ:DCSEL_CHEBY_GD}] using particular examples in~\figref{FIG:CASGD_AND_WFM}. (a)~$i=1$, $k=2,3,4$ in~\figref{FIG:GD_ODD} (all-odd coding set). (b)~$i=1$, $k=2,3,4$ in~\figref{FIG:GD_EVEN} (all-even coding set). }
    \label{FIG:GD_WFM_EXAMPLE}
\end{figure}

Figure~\ref{FIG:GD_WFM_EXAMPLE} provides graphical aids for qualitative study of the relationship between group delay profiles (including symmetry property) and the corresponding waveforms for the all-odd [\figref{FIG:GDWFMS_ODD}] and all-even [\figref{FIG:GDWFMS_EVEN}] sets. In each case, the left column plots three cascaded group delay profiles $\tau_{ik}(\omega)$, selected out from~\figref{FIG:CASGD_AND_WFM}, that contribute to the total MAI, $x_i(t)$. The right column plots the envelope waveform of the corresponding MAI contribution, $|h_{ik}(t)|$.

In general, it may be seen that group delay profile determines the distribution of the spectral contents (energy) of the signal in time. First, the group delay swing, $\Delta\tau_{ik}$, determines the time spread of $h_{ik}(t)$, which is widened by (sinc) ringing due to the chosen (rectangular) spectrum [Eq.~\eqref{EQ:DCMCONCEPT_CASTF}]. Second, the group delay minima and maxima regions of $\tau_{ik}(\omega)$, highlighted by grey boxes in the figure, determine the peak level of $h_{ik}(t)$. This fact may be best explained by an example. Consider the $(i,k)=(1,3)$ (middle) case in~\figref{FIG:GDWFMS_ODD}. Around the delay minima-maxima regions (left column), $\tau_{ik}'(\omega)\approx\delta\tau/\delta\omega$ is small, corresponding to relatively small $\delta\tau$ and large $\delta\omega$ ($\propto$ energy), and hence high energy concentration in a short time span ($\delta\tau$). A particular case is the linear group delay (no minima-maxima) [\figref{FIG:GDWFMS_ODD}, $(i,k)=(1,2)$], which leads to uniform energy distribution and flat amplitude level over the entire time span. Another particular case is when there are multiple subbands with the same delay [\figref{FIG:GDWFMS_EVEN}, $(i,k)=(1,3)$ and $(i,k)=(1,4)$], where the energies with different different carries but having identical delay times simply accumulate in the corresponding time span.

Note that, in practice, the transmitters are not synchronized and the channel delays are random, so that conditions 2) and 3) for the scenario leading to~\eqref{EQ:DCSEL_DECSIG} do not hold, which results in the partial MAI, $h_{ik}(t)$, being randomly shifted in time. This random time shifting would incur the worst case scenario for the total MAI, $x_i(t)$, when the peaks of all the contributions $h_{ik}(t)$ occur at the same time and add up constructively. The best way to minimize such a worst-case MAI peak is to minimize the peak of each $h_{ik}(t)$, which is achieved when the cascaded group delay profiles are
anti-symmetric [\figref{FIG:GDWFMS_ODD}] and hence equalize the energy distribution in both halves of the time span, whereas symmetric group delay profiles [\figref{FIG:GDWFMS_EVEN}] would instead tend to accumulate energy in half of the time span. This explains why the all-odd coding set is superior to the all-even coding set in terms of minimizing MAI peak level. Therefore, we will only use the all-odd coding set in the remainder of the paper.
%,

\subsection{Delay Swing -- Bandwidth Product}

Apart from the aforementioned MAI peak power (amplitude) due the energy concentration in time, the MAI average power is also of great concern since it is involved in the Signal-to-Interference Ratio~(SIR). We shall show that the MAI average power is in fact linearly proportional to the group delay swing and bandwidth product. For this purpose, we use the Parseval theorem~\cite{BK:2007_Andrei} to express the energy of the MAI $h_{ik}(t)$ as
\begin{equation}\label{EQ:DCSEL_PVL1}
  E  = \int_{-\infty}^{\infty}|h_{ik}(t)|^2\,dt = \int_{-\infty}^{\infty}|H_{ik}(2\pi f)|^2\,df,
\end{equation}
where $H_{ik}(2\pi f) = H_{ik}(\omega)$ is the cascaded transfer function [Eq.~\eqref{EQ:DCMCONCEPT_CASTF}]. Given the transfer magnitude (rectangular), Eq.~\eqref{EQ:DCSEL_PVL1} reduces
\begin{equation}\label{EQ:DCSEL_PVL2}
     \int_{-\infty}^{\infty} |h_{ik}(t)|^2\,dt = 2\int_{f_0-\Delta f/2}^{f_0 +\Delta f/2} 1 \,df =2\Delta f  = E,
\end{equation}
where the factor 2 includes the negative frequencies of $|H_{ik}(\omega)|$. The total MAI energy simply corresponds to $N-1$ identical contributions, i.e. $(N-1)E$. Then, assuming the total time MAI spread $2\Delta\tau$, neglecting the (sinc) ringing due to the finite (rectangular) spectrum, the average power of $x_i(t)$ is found as
\begin{equation}\label{EQ:DCSEL_MAIPOWTOT1}
    P_{\text{X}} \approx   \dfrac{(N-1)E}{2\Delta\tau} =  \dfrac{(N-1)\Delta f }{\Delta\tau },
\end{equation}
which is identical for all receivers, i.e. independent of $i$. Then, SIR corresponds to the peak power of $h_{ii}(t)$ [Eq.~\eqref{EQ:DCMCONCEPT_AUTOCORR}], which is $P_\text{S}=4\Delta f^2$, divided by $P_\text{X}$, leading to
\begin{equation}\label{EQ:DCSEL_SIR}
  \text{SIR} =  \dfrac{P_\text{S}}{P_{\text{X}}} = \dfrac{4\Delta\tau\Delta f }{N-1}.
\end{equation}
The term $\Delta\tau\Delta f$ is called Delay Swing-Bandwidth Product (DSBP), and Eq.~\eqref{EQ:DCSEL_SIR} clearly states that the SIR is linearly proportional to DSBP. Note that the signal peak power ($P_\text{S}$) is proportional to $\Delta f^2$, the MAI average power ($P_\text{X}$) is proportional to~$\Delta f/\Delta\tau$, increasing $\Delta f$ increases both $P_\text{S}$ and $P_\text{X}$ with $P_\text{S}$ increased faster, while increasing $\Delta\tau$ decreases only $P_\text{X}$, leading to the SIR linearly proportional to the DSBP.

\section{2X2 DCMA System Experimental Demonstration}\label{SEC:EXPERIMENT}

This section presents a proof-of-concept $2\times2$ DCMA system implementation, based on first-order (odd) Chebyshev coding, and demonstrates a corresponding data communication experiment in a LOS channel.

\subsection{System Implementation}

The system is designed to operate in the X-band, near which RF phaser technology has been best demonstrated so far, with the following specifications:

\begin{itemize}
  \item Bit rate: 200~Mb/s/channel (limited by our data generator), corresponding to bit period of 5~ns;
  \item Center frequency: $f_0=10$~GHz (X-band center);
  \item Pulse bandwidth: $\Delta f=4$~GHz from $8$ to $12$~GHz (X-band width).
\end{itemize}
Note that the spectral efficiency is not best explored here, we will characterize it in next section.

\begin{figure}[htbp]
\centering
    \includegraphics[width=1\columnwidth]{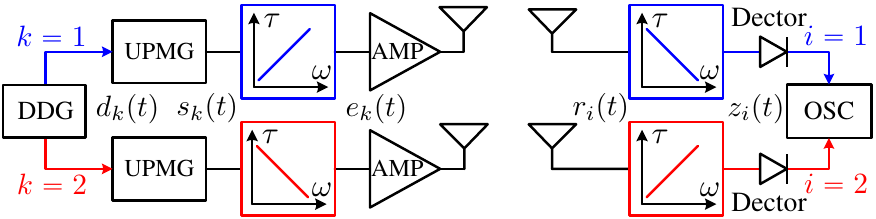}
%\psfragfig[width=1\linewidth, trim={0in 0in 0in 0in}]{sys2by2b}{
%\psfrag{D}[c][c][0.75]{DDG}
%\psfrag{A}[l][l][0.7]{AMP}
%\psfrag{O}[c][c][0.75]{OSC}
%\psfrag{M}[c][c][0.7]{UPMG}
%\psfrag{g}[l][l][0.9]{$\tau$}
%\psfrag{w}[r][r][0.9]{$\omega$}
%\psfrag{1}[l][l][0.85]{$d_k(t)$}
%\psfrag{2}[r][r][0.85]{$s_k(t)$}
%\psfrag{3}[l][l][0.85]{$e_k(t)$}
%\psfrag{4}[r][r][0.85]{$r_i(t)$}
%\psfrag{5}[r][r][0.85]{$z_i(t)$}
%\psfrag{6}[l][l][0.85]{\textcolor{blue}{$k=1$}}
%\psfrag{7}[l][l][0.85]{\textcolor{red}{$k=2$}}
%\psfrag{8}[r][r][0.85]{\textcolor{blue}{$i=1$}}
%\psfrag{9}[r][r][0.85]{\textcolor{red}{$i=2$}}
%\psfrag{t}[l][l][0.8]{Dector}
%}
\caption{Proposed system architecture for the proof-of-concept $2\times 2$ DCMA. ``DDG", ``UPMG", ``AMP" and ``OSC" stand for digital data generator,  UWB pulse modulator and generator, power amplifier and oscilloscope, respectively.}
\label{Fig:EXPERIMENT_2BY2_SYSTEM}
\end{figure}

The proposed prototype system architecture for the $2\times 2$ DCMA is shown in~\figref{Fig:EXPERIMENT_2BY2_SYSTEM}. The digital data generator produces two streams of baseband data, which are modulated and narrowed into UWB RF pulse trains by the UWB pulse modulator and generator, respectively, in each of the two transmitters ($k=1,2$). The corresponding UWB RF pulses are then encoded by respective phasers, amplified, and radiated by the antennas to the air, where they mix together. The two resulting signals, received by the two RX antennas, are decoded by the phasers, whose outputs are power-detected and finally displayed on the oscilloscope. The digital data generator is the Anritsu MP1630B digital analyzer, which is capable of generating baseband data with bit rate up to $200$~Mb/s. The oscilloscope is the Agilent DSO81204B digital oscilloscope, that captures RF signals with frequency up to 12~GHz. Since the system frequencies ($8 - 12$~GHz) are below the oscilloscope maximum measurable frequency, in the experiment we eliminate the power detector and, instead, simply square the acquired digital data stream (numerical power detection). At higher system operating frequencies, towards the millimeter-wave regime, direct UWB analog-to-digital conversion (ADC) may become an expensive or even not off-the-shelf solution. One should therefore use a real power detector. Amplification may be required depending on the communication distance. The other parts of the system will be introduced in the following sections.
 %The UPMG performs two functions: differential OOK modulation and UWB RF pulse generation. The DDG here is
%

\subsubsection{UWB Pulse Modulator and Generator (UPMG)}

\begin{figure}[h!t]
   \centering
\subfigure[]{\label{FIG:UPMGSCH}
    \includegraphics[width=0.8\columnwidth]{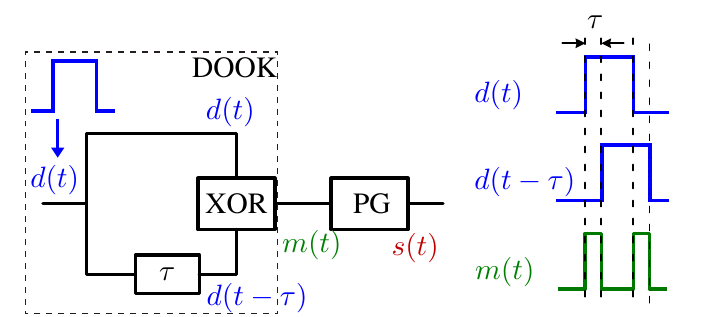}
%   \psfragfig[width=0.8\linewidth, trim={0in 0in 0in 0in}]{ooksch}{
%   \psfrag{a}[l][l][0.85]{\textcolor{blue}{$d(t)$}}
%   \psfrag{b}[l][l][0.85]{\textcolor{halfgreen}{$m(t)$}}
%   \psfrag{c}[l][l][0.85]{\textcolor{blue}{$d(t-\tau)$}}
%   \psfrag{X}[c][c][0.85]{XOR}
%   \psfrag{K}[r][r][0.85]{DOOK}
%   \psfrag{t}[c][c][0.85]{$\tau$}
%   \psfrag{x}[c][c][0.85]{\textcolor{purplered}{$s(t)$}}
%   \psfrag{R}[c][c][0.85]{PG}
%   }
}
\subfigure[]{\label{FIG:UPMGWFMS}
    \includegraphics[width=0.75\columnwidth]{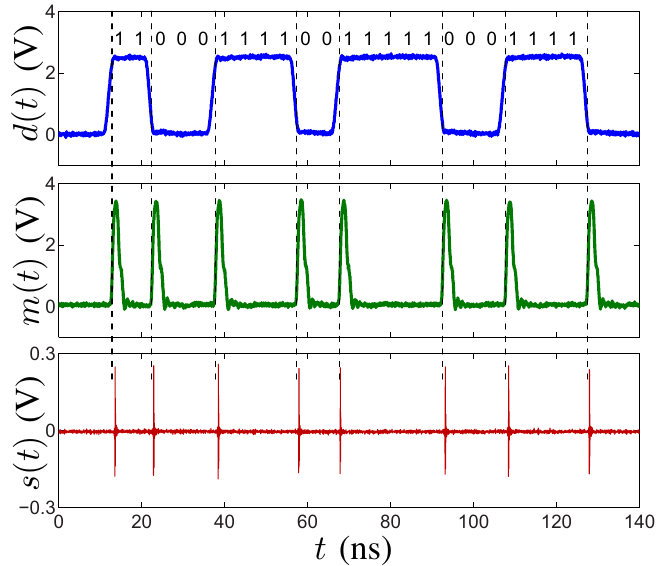}
%   \psfragfig[width=0.75\linewidth, trim={0in 0in 0in 0in}]{plsmgwfms}{
%   \psfrag{t}[c][c][1]{$t$~(ns)}
%   \psfrag{d}[ct][cm][1]{{$d(t)$~(V)}}
%   \psfrag{o}[ct][cm][1]{{$m(t)$~(V)}}
%   \psfrag{r}[ct][cm][1]{{$s(t)$~(V)}}
%   }
}
\subfigure[]{\label{FIG:UPWFMSPC}
    \includegraphics[width=1\columnwidth]{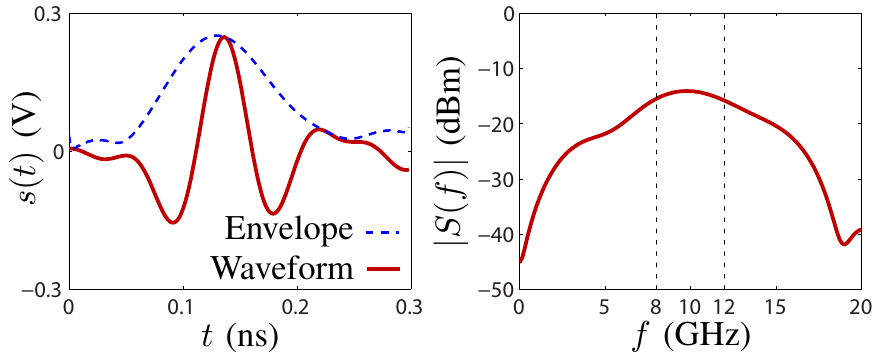}
%   \psfragfig[width=1\linewidth, trim={0in 0in 0in 0in}]{plswfmspc}{
%   \psfrag{t}[c][c][1]{$t$~(ns)}
%   \psfrag{f}[c][c][1]{$f$~(GHz)}
%   \psfrag{x}[ct][cb][1]{$s(t)$~(V)}
%   \psfrag{X}[c][c][1]{$|{S}(f)|$~(dBm)}
%   \psfrag{E}[r][r][1]{Envelope}
%   \psfrag{V}[r][r][1]{Waveform}
%   }
}
   \caption{UWB pulse modulator and generator (UPMG) schematic and corresponding waveform outputs. (a)~Schematic. (b)~Input data stream $d(t)$, corresponding DOOK modulation output $m(t)$ and RF pulse train $s(t)$. (c)~Zoom on single pulse and corresponding spectrum.}
   \label{FIG:UPMG}
\end{figure}

The UPMG performs two functions: Differential On-Off Keying (DOOK) modulation and UWB RF Pulse Generation (PG), as schematically shown in~\figref{FIG:UPMGSCH}. DOOK modulation is less used than non-differential On-Off Keying (OOK) modulation in practice. However, we adopt it here because the baseband data stream from the digital data generator is in the Non-Return-to-Zero (NRZ) format~\cite{BK:1999_JDGIBSON} and because the UWB RF pulse generator we use here is based on rising-edge triggering~\cite{JOUR:2016_ARXIV_LFZOU}. To properly trigger the pulse generator, OOK modulation would require to first transform NRZ into Return-to-Zero (RZ)~\cite{BK:1999_JDGIBSON} format, which requires iteratively detecting the level of the NRZ data every bit period, and hence requires a synchronized clocking circuit, whereas, DOOK modulation requires only detection of the rising and falling edges of the NRZ data, which may be easily done by using an XOR chip, as will be shown latter.

DOOK modulation produces mono-pulse when there is a change of state, i.e. rising or falling edge, occurring in the input data stream. This operation may be accomplished by a XOR chip. For detecting the change of state, the data stream $d(t)$ is divided into two paths, one of which is delay by $\tau$. The two signals, namely $d(t)$ and $d(t-\tau)$ are compared by the XOR chip, which generates 1 when the two inputs are different, and otherwise 0.  The width of the output mono-pulse $m(t)$ is equal to the delay, $\tau$. Here, as the bit duration is $5$~ns ($200$~Mb/s), $\tau$ should be less than $5$~ns, and we therefore set $\tau=3$~ns. Following the DOOK modulator is the UWB RF pulse generator (``PG") based on a pair of step recovery diodes~\cite{JOUR:2016_ARXIV_LFZOU}, which further narrows down the pulse width of $m(t)$ to the pico-second range. However, such narrowed UWB pulse $s(t)$ still contains a DC component, which cannot be radiated and is therefore eliminated by a reflection stub~\cite{JOUR:2001_TMTT_Lee}, yielding mono-cycle pulses, $s(t)$.

Figure~\ref{FIG:UPMGWFMS} shows the measured input data stream, $d(t)$, DOOK modulated pulses, $m(t)$, corresponding to the rising or falling edges of the data stream, and the UWB RF pulses, $s(t)$, respectively. Figure~\ref{FIG:UPWFMSPC} shows a zoom on a single RF pulse with its envelop and spectrum magnitude $|S(f)|$. The pulse waveform is not perfectly mono-cycle due to ringing~\cite{JOUR:2016_ARXIV_LFZOU}, and the envelope and spectrum profile are quasi-Gaussian. We may use a bandpass filter, not shown in~\figref{FIG:UPMGSCH}, to keep only the desired portion ($8-12$~GHz) of the spectrum so as to meet the system design specification. Within this bandwidth, the spectrum power density variation is found to be smaller than $2$~dB.

\subsubsection{LOS Wireless Channel}\label{SEC:EXPERIMENT:CHAN}

In the proof-of-concept system to be detailed next, we select a Vivaldi antenna for its wide bandwidth, high directivity and low dispersion~\cite{CONF:1979_EuMc_Gbison, CONF:2013_EMTS_Kikuta,JOUR:2009_PROCIEEE_Werner}. This antenna is practically well suited for directive LOS UWB applications. The Vivaldi antennas implemented here exhibit return loss higher than $10$~dB and broadside gain higher than $8$~dB and increasing with increased frequency over the bandwidth.

\begin{figure}[h!t]
   \centering
    \includegraphics[width=0.8\columnwidth]{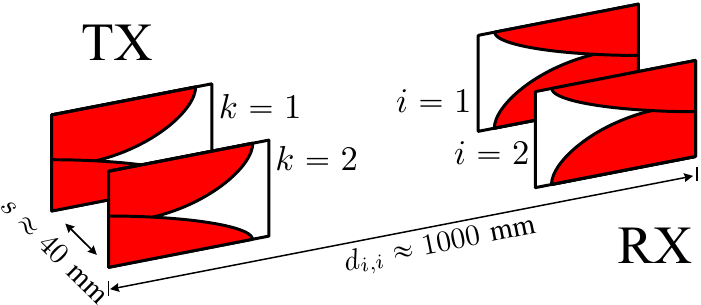}
%   \psfragfig[width=0.8\linewidth, trim={-0.4in -0.06in 0in 0in}]{chan2by2}{
%   \psfrag{1}[l][l][1]{$k=1$}
%   \psfrag{2}[l][l][1]{$k=2$}
%   \psfrag{3}[r][r][1]{$i=1$}
%   \psfrag{4}[r][r][1]{$i=2$}
%   \psfrag{s}[c][c][0.85]{$s\approx40$~mm}
%   \psfrag{d}[c][c][0.85]{$d_{ii}\approx1000$~mm}
%   }
   \caption{Vivaldi antenna setup for the measurement of the wireless channel responses.}
   \label{FIG:ANT_CONFIG}
\end{figure}

From the system point of view, the relevant channel response includes the antenna responses. For the LOS experiment to be performed here, four identical antennas are configured as shown in~\figref{FIG:ANT_CONFIG}. Two TX and two RX antennas are placed side by side with separation of $40$~mm, which is larger than one wavelength of the lowest frequency, $8$~GHz. The TX and RX antenna pairs are placed face to face, with distance $d_{ii}$ being approximately $1$~m, which satisfies the far-field condition. In this configuration, the $d_{ik} = \sqrt{d_{ii}^2+s^2}=1,001$~mm~$\approx d_{ii}=1,000$~mm, and the difference between communication distances $d_{ik}$ and $d_{ii}$ is only about $1$~mm, corresponding to $3$~ps delay difference. The corresponding reception angle difference is of only $2.3^\circ$, and the antenna gains may hence be assumed identical for all the communication links. Therefore, if two transmitters are sending signals at same time (synchronized) and same level, each receiver will receive the signal and the interference at the same time with indistinguishable magnitude before any further processing.

\begin{figure}[h!t]
   \centering
   \includegraphics[width=1\columnwidth]{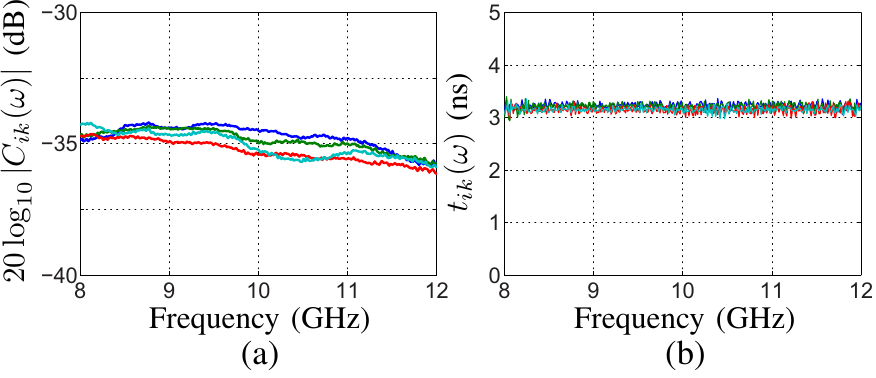}
%   \psfragfig[width=1 \linewidth, trim={0in -0.0in 0in 0in}]{measloschan}{
%   \psfrag{A}[c][c][1]{(a)}
%   \psfrag{B}[c][c][1]{(b)}
%  \psfrag{F}[c][c][0.9]{Frequency (GHz)}
%   \psfrag{M}[c][c][0.9]{$20\log_{10}|C_{ik}(\omega)|$~(dB)}
%   \psfrag{G}[c][c][0.9]{$t_{ik}(\omega)$~(ns)}
%   }
   \caption{Measured channel, corresponding to the configuration in~\figref{FIG:ANT_CONFIG}, using a vector network analyzer. (a)~Channel magnitudes, $|C_{ik}(\omega)|$, found to be approximately $-35$~dB with variation smaller $2$~dB. b)~Channel delays, $t_{ik}(\omega)$, found to be approximately $3.2$~ns.}
   \label{FIG:CHANNEL_RESPONSES}
\end{figure}

The LOS channel may be described by the transfer function $C_{ik}(\omega) = |C_{ik}(\omega)|\angle C_{ik}(\omega)$. For the LOS channel considered here, the magnitude response, $|C_{ik}(\omega)|$, includes free-space attenuation and antenna gain, as given by the Friis formula~\cite{BK:2011_MOLISCH_WIRELESSCOMM}. The former is inversely proportional to frequency, while the latter is proportional to frequency for a fixed antenna aperture, and hence the combination of the two yields a weakly frequency-dependent response. Moreover, the channel delay, $t_{ik}(\omega) = -\partial\angle C_{ik}/\partial\omega$, includes free-space delay, which is naturally non-dispersive, and antenna delay, which is weakly dispersive for  Vivaldi antennas~\cite{CONF:2013_EMTS_Kikuta,JOUR:2009_PROCIEEE_Werner}.
Figure~\ref{FIG:CHANNEL_RESPONSES} plots the measured channel magnitudes and delays, which are indeed essentially flat, indicating that both channel magnitudes and delays may be considered as weakly or negligibly frequency-dependent. Moreover, note that magnitudes and delays for the different channels are almost identical, due to the consideration of the previous paragraph.
%$\overline{c_\text{peak}}\approx4.536\times10^8$,

\subsubsection{Chebyshev Coding Phasers}

As mentioned in Sec.~\ref{SEC:DCSEL}, we choose first-order (odd) Chebyshev coding, corresponding to the coding set $\mathbf{C} = [1,-1]$, and set the group delay swing $\Delta\tau = 1$~ns. This group delay swing yields $\text{SIR}=16$ or $12$~dB according to~\eqref{EQ:DCSEL_SIR}.

\begin{figure}[h!t]
   \centering
      \includegraphics[width=0.75\columnwidth]{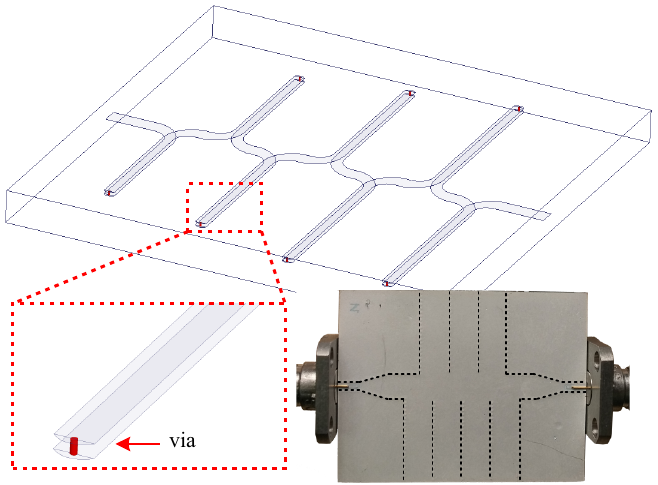}
   %\psfragfig[width=0.75\linewidth, trim={0in 0in 0in 0in}]{picphaser}{
   %\psfrag{V}[1][1][l]{via}
   %}
   \caption{Broadside-coupled C-section simulation layout (top) with shorted vias (bottom-left inset) and overall prototype (bottom-right inset).}
   \label{FIG:C-SECTION_PHASERS}
\end{figure}

\begin{figure}[h!t]
   \centering
         \includegraphics[width=1\columnwidth]{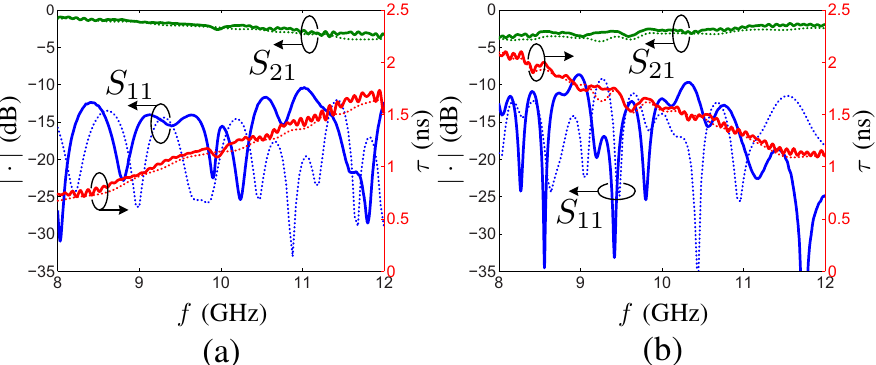}
   %\psfragfig[width=1 \linewidth, trim={0in -0.23in 0in 0in}]{phrsresp}{
%   \psfrag{A}[c][c][0.75]{$|\cdot|$~(dB)}
%   \psfrag{R}[r][r][1]{$S_{21}$}
%   \psfrag{T}[r][r][1]{$S_{11}$}
%   \psfrag{G}[c][c][0.75]{$\tau$~(ns)}
%   \psfrag{F}[c][c][0.75]{$f$~(GHz)}
%   \psfrag{a}[ct][ct][1]{(a)}
%   \psfrag{b}[ct][ct][1]{(b)}
%   }
   \caption{Measurement (solid) and simulation (dot) group delay ($\tau$) and magnitude responses ($S_{11}$ and $S_{21}$) for the phasers in~\figref{FIG:C-SECTION_PHASERS} with group delay swing $\Delta\tau=1$~ns. (a)~$+1^\text{th}$ order phaser. (b)~$-1^\text{th}$ order phaser.}
   \label{FIG:C-SECTION_PHASERS_RESULT1}
\end{figure}

The corresponding phasers are implemented in multilayered broadside-coupled C-section technology~\cite{JOUR:2010_TMTT_Gupta, JOUR:2012_MWCL_Horii} using a Low-Temperature Cofired-Ceramic (LTCC) fabrication process for high fabrication accuracy. Figure~\ref{FIG:C-SECTION_PHASERS} shows the fabricated circuit in the bottom right, with stripline-CPW-coaxial transition, and the grounding vias are highlighted. The inner structure is also shown by using the simulation layout, each broadside-coupled line pair is shorted at one end with via. Figure~\ref{FIG:C-SECTION_PHASERS_RESULT1} shows the measured results for the $+1$ and $-1$ order phasers.

\subsection{Measurement Results}

\begin{figure}
  \centering
\subfigure[]{\label{Fig:EXPERIMENT_2BY2_SYSTEM_TX}
\includegraphics[width=1\columnwidth]{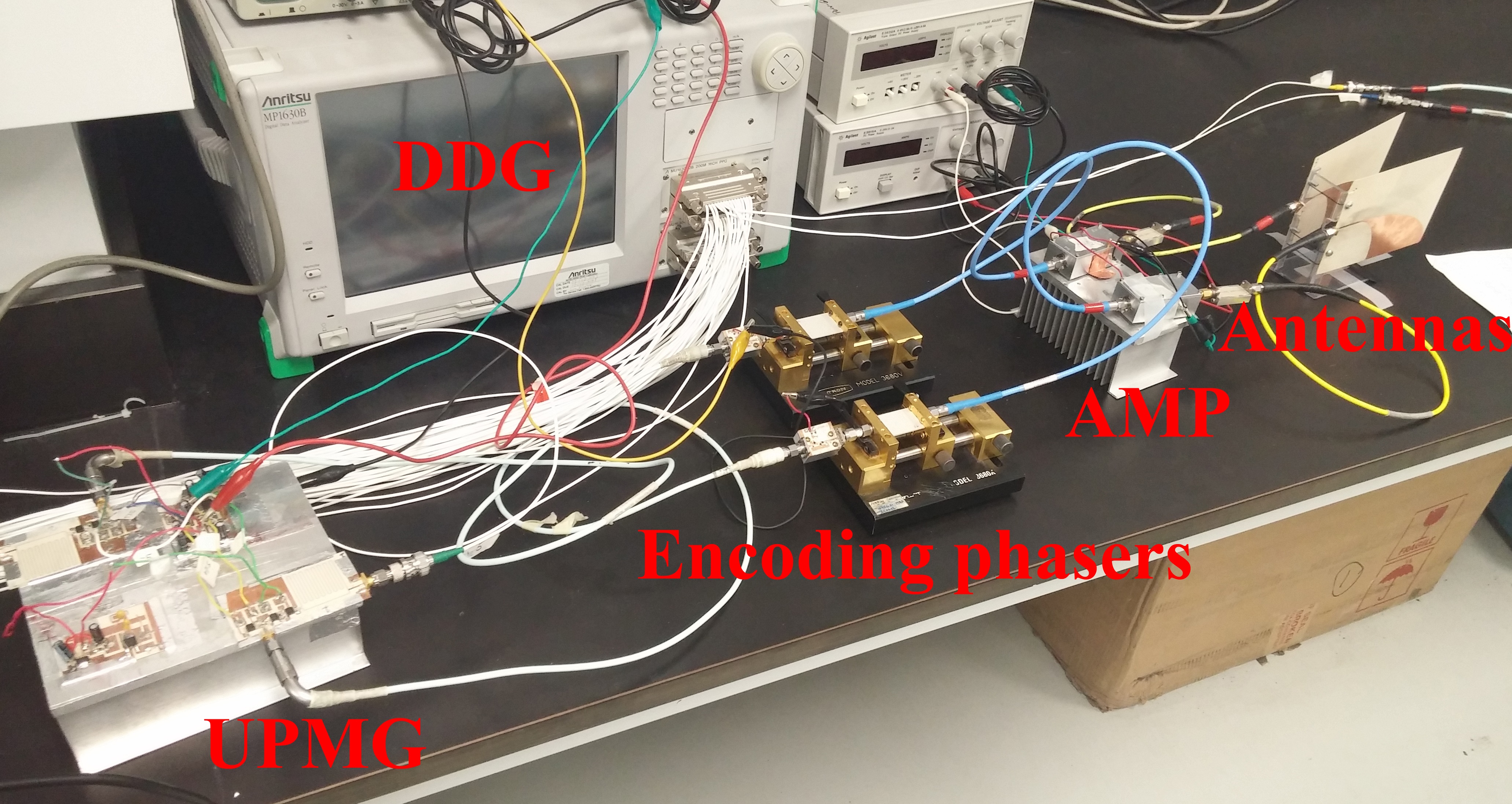}}
\subfigure[]{\label{Fig:EXPERIMENT_2BY2_SYSTEM_RX}
\includegraphics[width=1\columnwidth]{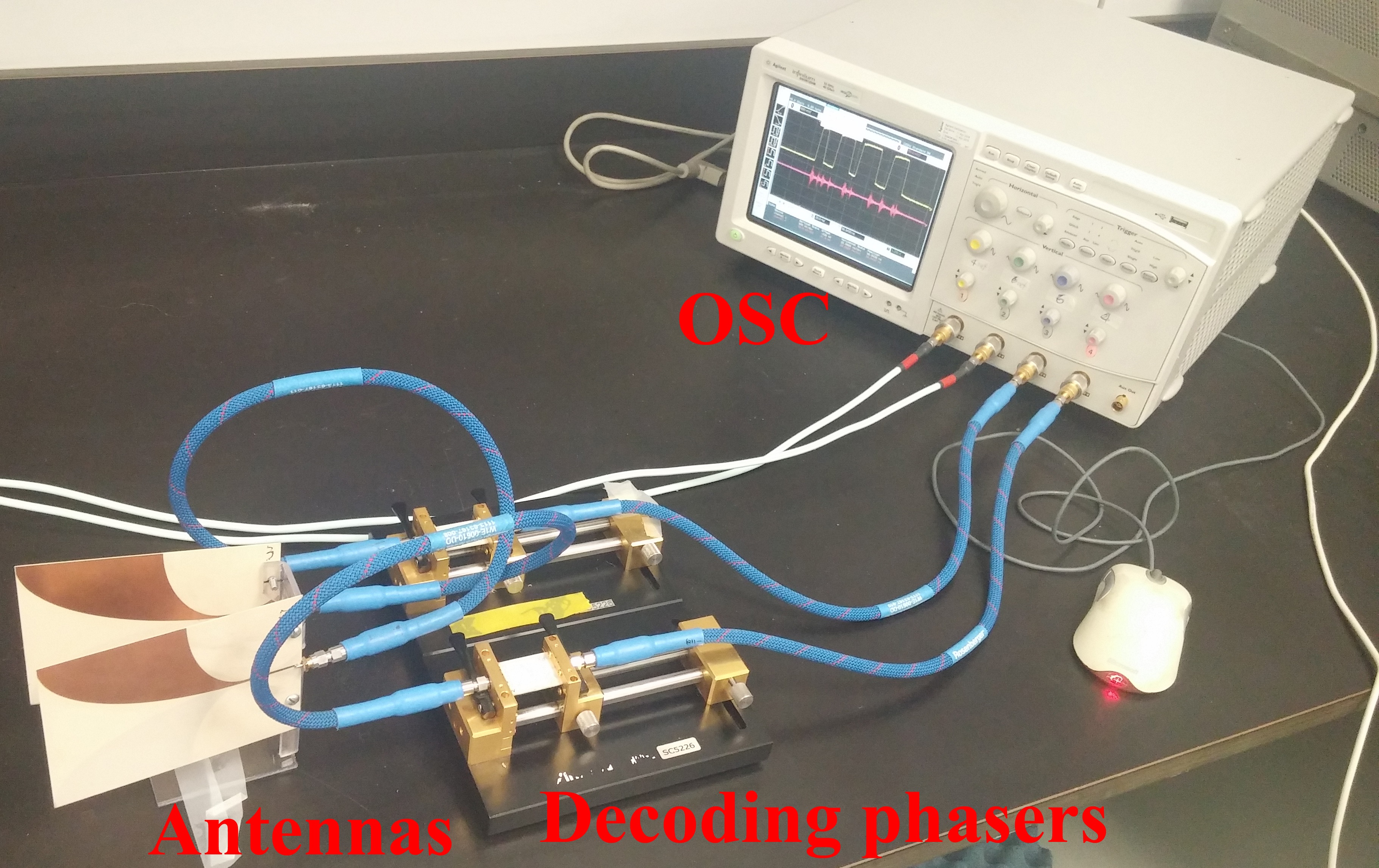}}
  \caption{Experimental prototype $2\times 2$ DCMA system corresponding to the architecture in~\figref{Fig:EXPERIMENT_2BY2_SYSTEM}. (a)~Transmitters. (b)~Receivers.}
  \label{FIG:PROTOTYPE_SYS}
\end{figure}

The experimental prototype system corresponding to the proposed architecture in~\figref{Fig:EXPERIMENT_2BY2_SYSTEM} is shown in~\figref{FIG:PROTOTYPE_SYS}. As already mentioned, the power detectors in~\figref{Fig:EXPERIMENT_2BY2_SYSTEM} are not included in this prototype system since power detection is to be done by taking square of the envelope of the waveform data captured by the oscilloscope. The bit period, which is the reverse of the bit rate, $T_\text{b}=1/R_\text{b}$, should be larger than the MAI spread, which here is $2\Delta\tau=2$~ns, to avoid intersymbol interference. Therefore, the theoretical maximal bit rate here would ideally approach $500$~Mb/s/channel given the $2$~ns MAI spread. Here we restrict the data rate to $200$~Mb/s/channel which corresponds to the maximal rate of the instrumentation.

\begin{figure}[h!t]
   \centering
   \includegraphics[width=1\columnwidth]{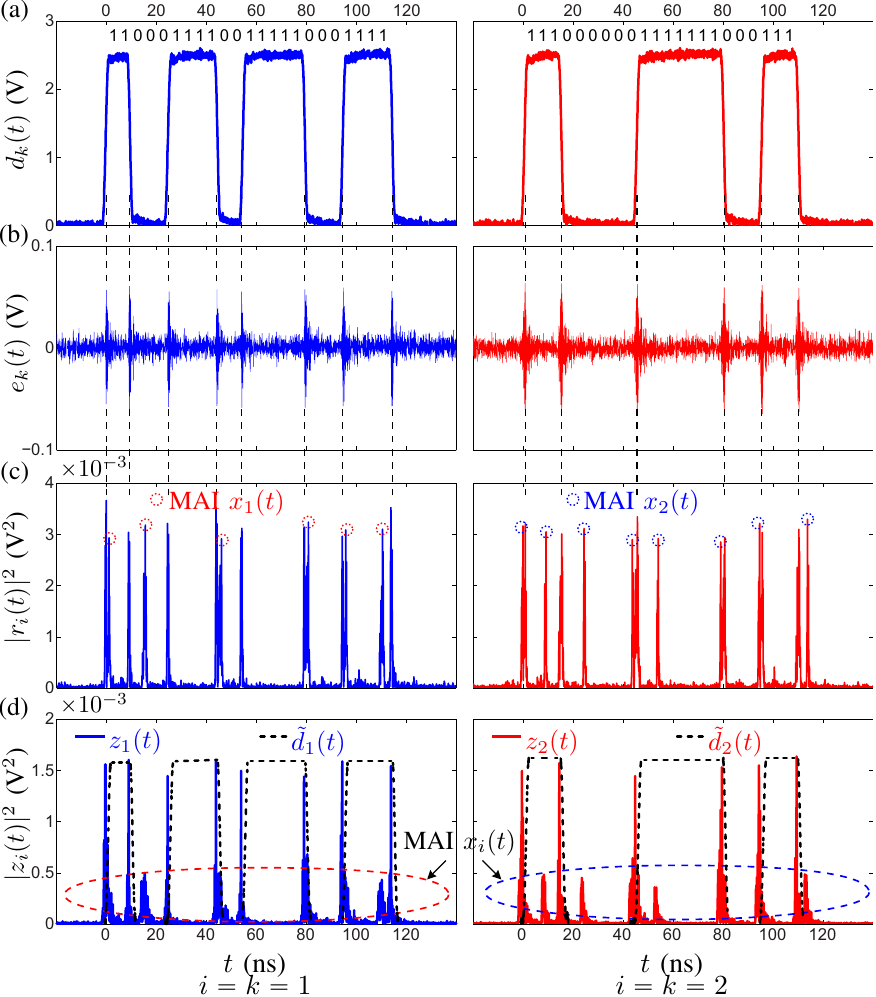}
%   \psfragfig[width=1 \linewidth, trim={-0.275in 0in 0in 0in}]{rxwfms}{
%   \psfrag{D}[b][b][0.75]{$d_k(t)$~(V)}
%   \psfrag{E}[b][b][0.75]{$e_k(t)$~(V)}
%   \psfrag{R}[b][b][0.75]{$|r_i(t)|^2$~(V$^2$)}
%   \psfrag{P}[b][b][0.75]{$|{z}_i(t)|^2$~(V$^2$)}
%   \psfrag{T}[ct][ct][0.8]{$t$~(ns)}
%   \psfrag{A}[ct][ct][0.8]{$i=k=1$}
%   \psfrag{B}[ct][ct][0.8]{$i=k=2$}
%   \psfrag{I}[l][l][0.75]{\textcolor{red}{MAI $x_1(t)$}}
%   \psfrag{N}[l][l][0.75]{\textcolor{blue}{MAI $x_2(t)$}}
%   \psfrag{m}[c][c][0.75]{MAI $x_i(t)$}
%   \psfrag{G}[l][l][0.75]{\textcolor{blue}{$\tilde{d}_1(t)$}}
%   \psfrag{H}[l][l][0.75]{\textcolor{red}{$\tilde{d}_2(t)$}}
%   \psfrag{C}[l][l][0.75]{\textcolor{blue}{$z_1(t)$}}
%   \psfrag{F}[l][l][0.75]{\textcolor{red}{$z_2(t)$}}
%   \psfrag{X}[r][r][0.75]{(a)}
%   \psfrag{V}[r][r][0.75]{(b)}
%   \psfrag{Y}[r][r][0.75]{(c)}
%   \psfrag{Z}[r][r][0.75]{(d)}
%   \psfrag{x}[l][l][0.7]{$\times10^{-3}$}
%   }
   \caption{Measured waveforms for the system in~\figref{FIG:PROTOTYPE_SYS} at the positions of the nodes indicated in~\figref{Fig:EXPERIMENT_2BY2_SYSTEM}. (a)~Baseband data stream, $d_k(t)$. (b)~Encoded waveforms, $e_k(t)$. (c)~Received waveform intensity, $|r_i(t)|^2$. (d)~Decoded waveform intensity, $|{z}_i(t)|^2$, and recovered baseband data, $\tilde{d}_i(t)$ by integration of $z_i(t)$.}
   \label{FIG:C-SECTION_PHASERS_RESULT}
\end{figure}

Figure~\ref{FIG:C-SECTION_PHASERS_RESULT} shows the system experimental results (waveforms) at different nodes in~\figref{Fig:EXPERIMENT_2BY2_SYSTEM}. Figure~\ref{FIG:C-SECTION_PHASERS_RESULT}(a) is the data stream, $d_k(t)$, from the digital data generator. The UWB RF pulses, $s_k(t)$, corresponding to the rising and falling edges of $d_k(t)$ based on DOOK modulation, have been shown in~\figref{FIG:UPMG}. Figure~\ref{FIG:C-SECTION_PHASERS_RESULT}(b) plots the signals, $e_k(t)$, encoded by the respective encoding phasers, which are dispersed versions of $s_k(t)$. Figure~\ref{FIG:C-SECTION_PHASERS_RESULT}(c) shows the received signal intensities, which correspond to the envelope squares, $|r_i(t)|^2$, and include the MAI from the undesired TX. The MAI peaks are circled out, and found to be roughly at the same level as the desired signals, and hence indistinguishable, at this stage. Note that the squared envelopes of $r_i(t)$ ($|r_i(t)|^2$) here do not correspond to the actual operation of the system at the corresponding nodes in~\figref{Fig:EXPERIMENT_2BY2_SYSTEM}; we numerically perform this operation for comparison with the decoded and power detected outputs, $|z_i(t)|^2$, as will be shown in~\figref{FIG:C-SECTION_PHASERS_RESULT}(d). Figure~\ref{FIG:C-SECTION_PHASERS_RESULT}(d) plots the intensity of the signals, ${z}_i(t)$ [$|{z}_i(t)|^2$] decoded by the respective decoding phasers. The MAI levels are dramatically reduced compared to those in~\figref{FIG:C-SECTION_PHASERS_RESULT}(c), hence allowing to correctly detect the desired information signal. Finally, one may use integration to recover the baseband data stream, $\tilde{d}_i(t)$, not physically performed in the experiment but shown in the black dashed lines in~\figref{FIG:C-SECTION_PHASERS_RESULT}(d). The power ratio of the signal peak to the MAI Peak-Envelope-Power (PEP) ranges from $5$ to $6$. One may find the MAI average power from the PEP and then further obtain the SIR. Ideally, the MAI is a flat-amplitude curve due to the linear group delay, according to the analysis in~\secref{SEC:DCSEL}. The MAI average power is typically half of its PEP, and  the SIR should then be twice of the power ratio of the signal peak to the MAI PEP, which is about $10$ to $12$, in reasonable agreement with the value $\text{SIR}=16$ obtained by~\eqref{EQ:DCSEL_SIR}, given the various non-idealities involved in the experiment.

\section{LOS Wireless System Characterization}\label{SEC:SYS}

The previous section experimentally demonstrated a proof-of-concept $2\times 2$ DCMA system with identical LOS channels [$c_{ik}(t)=\delta(t)$, $\forall i,k$] and without any specific noise study. This section will characterize the performance of a general DCMA system comprised of $N\times N$ TX-RX user pairs with different and arbitrary LOS channels [$c_{ik}(t)=a_{ik}\delta(t-t_{ik})$, with random $a_{ik}$ and $t_{ik}$] and in terms of noise. This characterization will be performed analytically and numerically, as it would be excessively involved to perform experimentally, but it will provide the key results relevant to the design of future complex DCMA systems. For simplicity, we shall assume that the environment is time-invariant (static) and open-boundary, i.e. without echos.

\subsection{Channel Description}\label{SEC:DCMACHAN}

The LOS channel impulse response may be found by taking the inverse Fourier transform of the transfer function, $C_{ik}(\omega)$, which implicitly accounts for the antennas and may be obtained by measurement, as in~\secref{SEC:EXPERIMENT:CHAN}, or by analytical formulas. For the case of a free-space LOS channel, the application of Friis formula with added delay leads to~\cite{BK:2011_MOLISCH_WIRELESSCOMM}
\begin{equation}\label{EQ:LOSCHANTF_FRIIS}
\begin{split}
   C_{ik}(\omega) &= \dfrac{\lambda\sqrt{G_{\text{TX}_k}(\omega)G_{\text{RX}_i}(\omega)}}{4\pi d_{ik}}e^{-j\omega t_{ik}}e^{-j\int_{-\infty}^{\omega}t'_{ik}(\omega')\,d\omega'}\\
    & = \dfrac{\sqrt{G_{\text{TX}_k}(\omega)G_{\text{RX}_i}(\omega)}}{2\omega t_{ik}}e^{-j\omega t_{ik}}e^{-j\int_{-\infty}^{\omega}t'_{ik}(\omega')\,d\omega'},
\end{split}
\end{equation}
where $G_{\text{TX}_k}(\omega)$ and $G_{\text{RX}_i}(\omega)$ are the power gain functions in the TX$_k$ and RX$_i$ directions, respectively, $t_{ik}=d_{ik}/c$ is the free-space (non-dispersive) delay, $c$ is the speed of light, and $t'_{ik}(\omega)$ is the channel dispersive delay induced by the antennas. In a practical UWB pulse (as opposed to narrow multi-band) system, the channel response is essentially frequency independent in both delay, i.e. $\partial t'_{ik}/\partial\omega\approx0$, and in magnitude, i.e. $|C_{ik}(\omega)|\approx {\sqrt{G_{\text{TX}_k}(\omega_0)G_{\text{RX}_i}(\omega_0)}}/{2\omega_0t_{ik}}$, where $\omega_0$ is the center frequencym as illustrated in \figref{FIG:CHANNEL_RESPONSES}. Therefore, denoting the product of the TX$_k$ and RX$_i$ antenna power gains $G_{ik}=G_{\text{TX}_k}(\omega_0)G_{\text{RX}_i}(\omega_0)$ and shifting the reference time to $t'_{ik}(\omega) = 0$, Eq.~\eqref{EQ:LOSCHANTF_FRIIS} reduces to the expression
\begin{equation}\label{EQ:LOSCHANTF_FRIIS1}
  C_{ik}(\omega) = \dfrac{\sqrt{G_{ik}}}{2\omega_0t_{ik}}e^{-j\omega t_{ik}},
\end{equation}
whose impulse response is
\begin{equation}\label{EQ:LOSCHANIR_FRIIS}
  c_{ik}(t)  = \dfrac{\sqrt{G_{ik}}}{2\omega_0t_{ik}}\delta(t-t_{ik}).
\end{equation}

Since the transmission throughput is ultimately determined by SIR, which includes the ratio $a_{ii}/a_{ik}$, we further normalize~\eqref{EQ:LOSCHANIR_FRIIS} as
\begin{subequations}\label{EQ:LOSCHAN_IR0}
 \begin{equation}\label{EQ:LOSCHAN_IR0_IK}
 \begin{split}
     c_{ik}'(t) &= \dfrac{a_{ik}}{a_{ii}}\delta(t-t_{ik})\\
  & = \dfrac{\sqrt{G_{ik}/G_{ii}}}{t_{ik}/t_{ii}}\delta(t-t_{ik}),
 \end{split}
\end{equation}
which finally yields
 \begin{equation}\label{EQ:LOSCHAN_IR0_II}
  c_{ii}'(t) = \delta(t-t_{ii}).
\end{equation}
\end{subequations}

\subsection{MAI Analysis}\label{SEC:MAICHAR}

In order to fully characterize the system in the $c_{ik}(t)$ channel described by~\eqref{EQ:LOSCHAN_IR0}, we shall first determine its MAI in that channel. For simplicity, we idealize the information signal to a train of OOK-modulated Dirac pulses with bit period $T_\text{b}=2\Delta\tau$, corresponding to the maximal dispersed duration, i.e.
\begin{equation}\label{EQ:SYS_SIG}
\begin{split}
 &s_k(t)=\sum_{\ell=1}^{\infty}d_{k,\ell}g_k\delta(t-{\ell}T_\text{b}-t_{\text{TX}_k}),\\
 &k\in\{1,\dots,i,\dots,N\},\quad g_i=1,
\end{split}
\end{equation}
where $d_{k,\ell}=1$ or $0$ is the $\ell^\text{th}$ bit from TX$_k$, $g_k>0$ is the normalized transmitted signal amplitude, and $ t_{\text{TX}_k}$ is a random transmitting delay taking account for the transmitter asynchronization.
Inserting~\eqref{EQ:LOSCHAN_IR0_II} and~\eqref{EQ:LOSCHAN_IR0_IK} into~\eqref{EQ:DCMCONCEPT_DECSIGWFM_DESIRED} and \eqref{EQ:DCMCONCEPT_DECSIGWFM_MAI}, respectively, with~\eqref{EQ:SYS_SIG}, yields
\begin{subequations}\label{EQ:ASYNC_DECSIGS}
\begin{equation}\label{EQ:ASYNC_DECSIG_DESIRED}
  \tilde{s}_i(t) = \sum_{\ell=1}^{\infty}d_{i,\ell} \alpha_{ii}h_{ii}(t-{\ell}T_\text{b}-t_{\text{TX}_i}-t_{ii}),
\end{equation}
\begin{equation}\label{EQ:ASYNC_DECSIG_MAI}
\begin{split}
    x_i(t) =\sum_{\ell=1}^{\infty} \sum_{\substack{k=1\\k\neq i}}^{N}d_{k,\ell}\alpha_{ik} h_{ik}(t-{\ell}T_\text{b}-t_{\text{TX}_k}-t_{ik}).
\end{split}
\end{equation}
where the overall channel amplitude,
\begin{equation}\label{EQ:ASYNC_DECSIG_AMP}
  \alpha_{ik} = g_k\dfrac{\sqrt{G_{ik}/G_{ii}}}{t_{ik}/t_{ii}} \quad (\alpha_{ik}>0)
\end{equation}
\end{subequations}
is random. This quantity has been normalized for the desired signal magnitude to be unity, i.e. $\alpha_{ii}\equiv 1$ in~\eqref{EQ:ASYNC_DECSIG_DESIRED}, while we would have $\alpha_{ik}<1$, $\alpha_{ik}=1$ and $\alpha_{ik}>1$ ($k\neq i$) in~\eqref{EQ:ASYNC_DECSIG_MAI} for the cases of lossy, lossless and gain MAI signals, respectively.

Then, the total decoded signal, which is according to~\eqref{EQ:DCMCONCEPT_DECSIGWFM} the sum of~\eqref{EQ:ASYNC_DECSIG_DESIRED}, \eqref{EQ:ASYNC_DECSIG_MAI} and $n(t)$, becomes
\begin{equation}\label{EQ:ASYNC_DECSIG}
\begin{split}
    z_i(t)
    =&  \sum_{\ell=1}^{\infty}d_{i,\ell} h_{ii}(t-{\ell}T_\text{b}-t_{\text{TX}_i}-t_{ii})\\
    & +\sum_{\ell=1}^{\infty} \sum_{\substack{k=1\\k\neq i}}^{N}d_{k,\ell}\alpha_{ik}  h_{ik}(t-{\ell}T_\text{b}-t_{\text{TX}_k}-t_{ik})\\
     &+n(t),
\end{split}
\end{equation}

Note that the channel delay, $t_{ik}$, MAI channel amplitude, $\alpha_{ik}$ ($k\neq i$), and transmitting delay, $t_{\text{TX}_k}$, are all random variables. The channel delay is a uniform random variable, corresponding to
\begin{equation}\label{EQ:CHAN_DELAY_DIST}
  t_{ik}   =\mathcal{U}\left(d_\text{min}/c, d_\text{max}/c\right),
\end{equation}
where $\mathcal{U}(\cdot)$ denotes the uniform  distribution (or rect function) with support $[d_\text{min}/c, d_\text{max}/c]$, and $d_\text{min}$ and $ d_\text{max}$ are the minimum and maximum communication distances, and are respectively set to range from 0 to  $4$~meters in later characterization.
The overall channel amplitude, while being affected by the channel delay ratio $t_{ik}/t_{ii}$, which is fixed in static LOS wireless environment, depends on the antenna gain ratio, $\sqrt{G_{ik}/G_{ii}}$, and the transmitting signal amplitude level, $g_k$. $\sqrt{G_{ik}/G_{ii}}$ and $g_k$ may be tuned using beam forming and power amplification, respectively. Therefore, one may assume $\alpha_{ik}$ to be independent of $t_{ik}/t_{ii}$ and thus $\alpha_{ik}^2$ (intensity) to globally also follow a uniform distribution, i.e.
\begin{equation}\label{EQ:CHAN_ENGY_DIST}
 \alpha_{ik}^2  =\mathcal{U}\left(\alpha_\text{min}^2, \alpha_\text{max}^2\right).
\end{equation}
In particular, $\alpha_{ik}^2\equiv1$ for $\alpha_\text{min}^2=\alpha_\text{max}^2=1$, which corresponds to the case where all the channels have equal energy. Finally, the transmitting delay meaningfully varies within $[0, T_\text{b}]$, given that $\tilde{s}_{ii}(t)$ and $x_i(t)$ in~\eqref{EQ:ASYNC_DECSIGS} are both periodic functions with period $T_\text{b}=2\Delta\tau$. Thus,
\begin{equation}\label{EQ:TX_DELAY_DIST}
  t_{\text{TX}_k}   =\mathcal{U}\left(0, T_\text{b}\right),~\text{with}~T_\text{b} = 2\Delta\tau.
\end{equation}

One may next determine the statistical distribution of the MAI, $x_i(t)$ [Eq.~\eqref{EQ:ASYNC_DECSIG_MAI}], which is described by the corresponding probability density function (PDF). The analytical derivation of the PDF of MAI is beyond the scope of this paper, and  readers are referred to specialized works in this area for details.  For instance, the authors of~\cite{JOUR:1999_JLT_Fathallah} use the normal distribution to approximate the MAI distribution in  Bragg grating fiber based frequency hopping multiple access, while the authors of~\cite{JOUR:2005_TSP_SGZICI} analytically show that MAI asymptotically approaches the normal distribution with increased frames per symbol in time-hopping multiple access. Here, we will numerically show that the MAI distribution in Chebyshev coding DCMA may also be approximated by the normal distribution.

For this purpose, we use the random variable $x_i$ as the time sample of the MAI, so that
 \begin{subequations}\label{EQ:ASYNC_MAISAMPLES_RXI}
 \begin{equation}\label{EQ:ASYNC_MAISAMPLES_RXI1}
   x_i\in\pmb{x}_i =\{ x_{i,1},\ldots, x_{i,p}, \ldots, , x_{i,L}\},
    \end{equation}
where
\begin{equation}
   x_{i,p} =  x_i(p t_\text{s}),\quad p\in\mathbb{Z},
\end{equation}
 \end{subequations}
where $t_\text{s}$ is sampling period, $L=2T_\text{b}/t_\text{s}$ is the sample length, and $x_{i,p}$ is the time sample in~\eqref{EQ:ASYNC_DECSIG_MAI} at $pt_\text{s}$. At this point, we statistically compute the corresponding mean ($\mu_{\text{X},i}$) and standard deviation ($\sigma_{\text{X},i}$) of $x_i$ as
\begin{equation}\label{EQ:ASYNC_MEAN_STD_RXI}
     \mu_{\text{X},i} = \dfrac{1}{L_i}\sum_{p=1}^{L_i} x_{i,p}, \quad \sigma_{\text{X},i}^2 = \dfrac{1}{L_i}\sum_{p=1}^{L_i}( x_{i,p}-\mu_{\text{X},i})^2,
\end{equation}
with $\sigma_{\text{X},i}^2$ being the average power of MAI ${x}_i(t)$. For the purpose of plotting the normalized PDF, one first normalizes the random variable $x_i$ and its mean, $\mu_{\text{X},i}$, and standard deviation, $\sigma_{\text{X},i}$, to the peak value of the desired signal~[Eq.~\eqref{EQ:DCMCONCEPT_AUTOCORR}], i.e.
\begin{equation}\label{EQ:RVNORM}
  \hat x_i = \dfrac{x_i}{2\Delta f},\quad
  \hat \mu_{\text{X},i} = \dfrac{\mu_{\text{X},i}}{2\Delta f},\quad
  \hat \sigma_{\text{X},i}^2 = \dfrac{\sigma_{\text{X},i}^2}{4\Delta f^2},
\end{equation}
which provides all the parameters of the normal distribution PDF of $\hat x_i$,
 \begin{equation}\label{EQ:ASYNC_GPDF}
   \text{PDF}_{\text{X},i}(\hat x_i) = \dfrac{1}{\sqrt{2\pi\hat \sigma_{\text{X},i}^2}}\exp\left[-\dfrac{(\hat x_i- \hat \mu_{\text{X},i} )^2}{2\hat \sigma_{\text{X},i}^2}\right],
 \end{equation}
while the PDF could also be obtained by counting the occurrences of the different values of $\hat x_i$.

Note that the normalized MAI variance, $\hat\sigma_{\text{X},i}^2$, is the interference to signal power ratio, which is exactly the opposite of SIR, i.e.
\begin{equation}\label{EQ:SIR}
 \text{SIR}_i= \dfrac{1}{\hat \sigma_{\text{X},i}^2}.%=\dfrac{4\Delta f^2}{\sigma_{\text{X},i}^2}
\end{equation}

In~\secref{SEC:DCSEL}, we have derived the MAI power and SIR for the case of equal channel energy ($\alpha_{i,k}=1,\,\forall\,k$) based on the procedure from~\eqref{EQ:DCSEL_PVL1} to~\eqref{EQ:DCSEL_SIR}, so as to find the relationship between the DSBP (Delay Swing-Bandwidth Product) and SIR. We shall now extend this relationship to cover the case  where the channels have different energies, i.e. $\alpha_{ik_1}\neq\alpha_{ik_2}$. Note that each $h_{ik}(t)$ in~\eqref{EQ:ASYNC_DECSIG_MAI} has a factor $\alpha_{ik}$ and hence energy $\alpha_{ik}^2E$, which leads to the average power of $x_i(t)$
\begin{subequations}\label{EQ:MAIPOW_GENERAL}
\begin{equation}\label{EQ:MAIPOW_GENERAL1}
  P_{\text{X},i} = \dfrac{\sum_{\substack{k=1\\k\neq i}}^{N}\alpha_{ik}^2E}{2\Delta\tau}
  =\alpha_i^2\dfrac{(N-1)\Delta f}{\Delta\tau},
\end{equation}
where
\begin{equation}\label{EQ:MAIPOW_GENERAL2}
  \alpha_i^2= \dfrac{1}{N-1}\sum_{\substack{k=1\\k\neq i}}^{N}\alpha_{ik}^2
\end{equation}
\end{subequations}
is the arithmetic mean of $\alpha_{ik}^2$ [Eq.~\eqref{EQ:CHAN_ENGY_DIST}] and $E=2\Delta f$ is the energy of $h_{ik}(t)$ over one bit period ($T_\text{b}=2\Delta\tau$). The corresponding SIR is then found as
\begin{equation}\label{EQ:SIR2}
  \text{SIR}'_i = \dfrac{4\Delta f^2}{P_{\text{X},i}} = \dfrac{4\Delta f\Delta\tau}{\alpha_i^2(N-1)}
  =\dfrac{4\cdot\text{DSBP}}{\alpha_i^2(N-1)}.
\end{equation}
We will later compare SIR$_i'$ [Eq.~\eqref{EQ:SIR2}] and $P_{\text{X},i}$ [Eq.~\eqref{EQ:MAIPOW_GENERAL1}] to SIR$_i$ [Eq.~\eqref{EQ:SIR}] and $\sigma_{\text{X},i}^2$ [Eq.~\eqref{EQ:ASYNC_MEAN_STD_RXI}], respectively, and show that the former are good estimates for the latter, which avoids resorting to the statistical results~\eqref{EQ:SIR} and~\eqref{EQ:ASYNC_MEAN_STD_RXI}.

\begin{figure}[h!t]
   \centering
   \includegraphics[width=1\columnwidth]{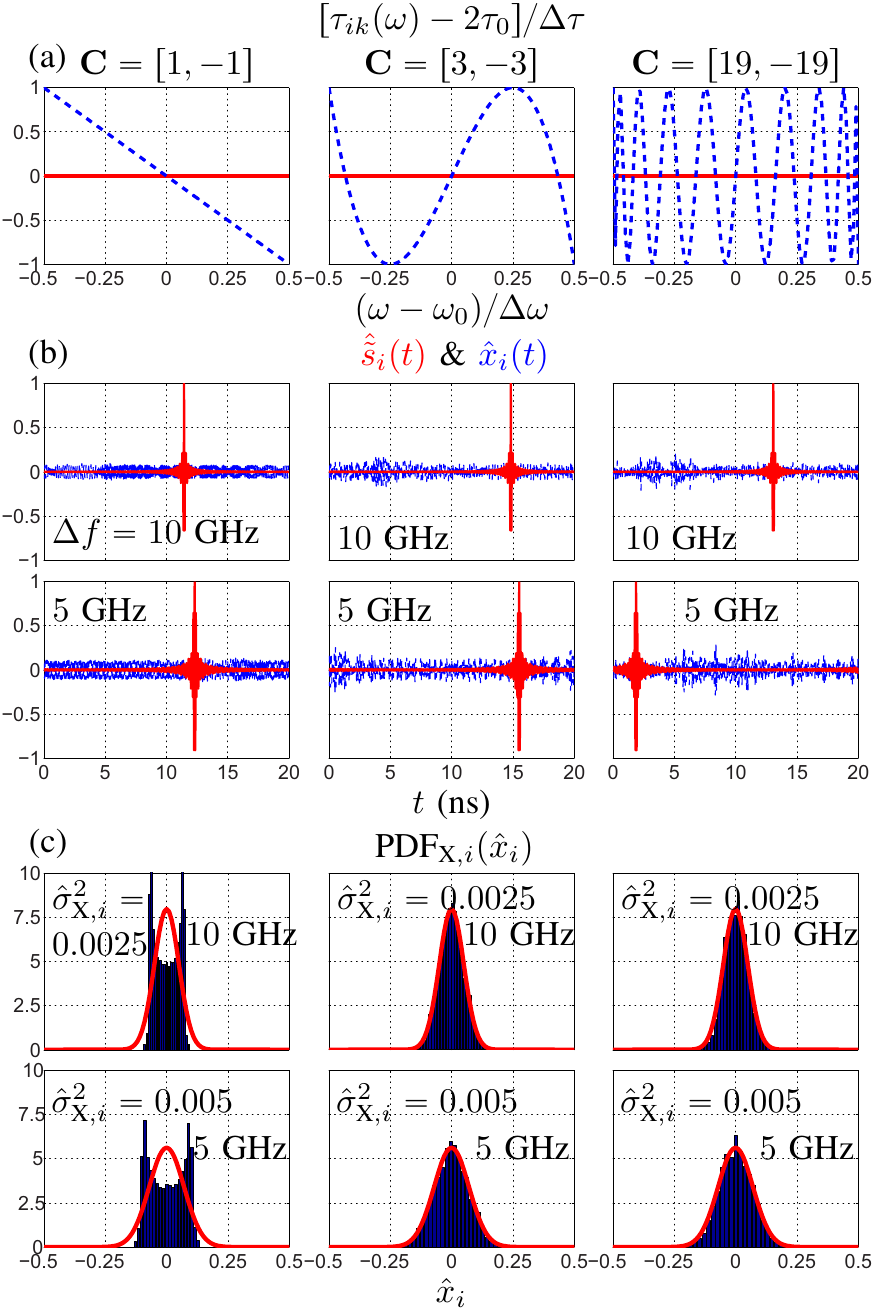}
%   \psfragfig[width=1\linewidth, trim={0in 0in  0in -0.4in}]{maipdf}{
%   \psfrag{G}[cb][ct][1]{${[\tau_{ik}(\omega)-2\tau_0]}/{\Delta\tau}$}
%   \psfrag{S}[cb][cb][1]{\textcolor{red}{${\hat{\tilde{s}}_i(t)}$} \& \textcolor{blue}{${\hat{x}_i(t)}$}}
%   \psfrag{P}[cb][cb][1]{$\text{PDF}_{\text{X},i}(\hat x_i)$}
%   \psfrag{F}[c][c][1]{$(\omega-\omega_0)/\Delta\omega$}
%   \psfrag{T}[c][c][1]{$t$~(ns)}
%   \psfrag{X}[c][c][1]{$\hat x_i$}
%   \psfrag{A}[c][l][1]{(a)}
%   \psfrag{B}[c][l][1]{(b)}
%   \psfrag{C}[c][l][1]{(c)}
%   \psfrag{D}[cb][cb][1]{$\mathbf{C}=[1,-1]$}
%   \psfrag{E}[cb][cb][1]{$\mathbf{C}=[3,-3]$}
%   \psfrag{H}[cb][cb][1]{$\mathbf{C}=[19,-19]$}
%   \psfrag{K}[lt][lt][1]{$\hat\sigma_{\text{X},i}^2=$}
%   \psfrag{M}[lt][lt][1]{$0.0025$}
%   \psfrag{U}[lt][lt][1]{$\hat\sigma_{\text{X},i}^2=0.0025$}
%   \psfrag{V}[lt][lt][1]{$\hat\sigma_{\text{X},i}^2=0.005$}
%   \psfrag{O}[lb][lb][1]{$\Delta f=10$~GHz}
%   \psfrag{R}[lb][lb][1]{$10$~GHz}
%   \psfrag{N}[lt][lt][1]{$5$~GHz}
% }
   \caption{Three $2\times2$ DCMA examples (corresponding to the three columns) with $\sigma_\text{N}=0$ (noiseless), $\alpha_{ik}^2\equiv1\,(\forall\,i,k)$ [Eq.~\eqref{EQ:CHAN_ENGY_DIST} with $\alpha_\text{min}^2=\alpha_\text{max}^2=1$]. (a)~Code sets $\mathbf{C}$ and corresponding cascaded group delays $\tau_{ik}(\omega)$ [Eq.~\eqref{EQ:DCMCONCEPT_CASGD} with~\eqref{EQ:DCSEL_CHEBY_GD} and $\Delta\tau=10$~ns] with $(i,k)=(1,1)$ (red solid) and $(i,k)=(1,2)$ (blue dashed). (b)~Corresponding waveforms over one bit, $T_\text{b}=2\Delta\tau = 20$~ns: desired signal $\hat{\tilde{s}}_i(t)$ (red solid) and MAI $\hat{x}_i(t)$ (blue dashed) [Eq.~\eqref{EQ:ASYNC_DECSIGS} normalized by $2\Delta f$] for $\Delta f = 10$~GHz and $\Delta f = 5$~GHz cases. (c)~Corresponding actual MAI distribution (blue bars) and approximated normal distribution [Eq.~\eqref{EQ:ASYNC_GPDF}] with $\hat\sigma_{\text{X},i}^2$ values [interference to signal ratio: Eq.~\eqref{EQ:RVNORM}] for $\Delta f = 10$~GHz and $\Delta f = 5$~GHz cases.}
   \label{FIG:DECSIG_EXAMPLE}
\end{figure}

Figure~\ref{FIG:DECSIG_EXAMPLE} shows three examples of $2\times 2$ DCMA coding sets and their corresponding cascaded group delays, waveforms and MAI distributions in a noiseless channel with identical energies ($\alpha_{ik}^2\equiv1$). In these examples, only one MAI contribution exists. For the coding order $m_i\neq\pm1$ [Eq.~\eqref{EQ:DCSEL_CHEBY_CODESET}], the actual MAI distribution agrees well with the normal distribution given by~\eqref{EQ:ASYNC_GPDF} [\figref{FIG:DECSIG_EXAMPLE}(c) second and third columns]. However, when $m_i=\pm 1$, the approximation of normal distribution does not hold [\figref{FIG:DECSIG_EXAMPLE}(c) first column]\footnote{The MAI distribution for a linear group delay exhibits, instead of a normal distribution, a pair sharp rays at the polar value corresponding to the quasi-uniform value of the waveform [see for instance \figref{FIG:GDWFMS_ODD} for the $(i,k)=(1,2)$], and decreasing magnitude towards zero, corresponding to faster transitions between positive and negative signal values.}. Therefore, in the following characterization, we will exclude the $m_i=\pm1$ code.

The MAI distributions are found to be always symmetric about $\hat x_i=0$, i.e. $\mu_{\text{X},i}\equiv 0$, which is easily explained from the fast polar oscillation of $\hat{x}_i(t)$ about zero and the consequent global cancellation in the corresponding average process in~\eqref{EQ:ASYNC_MEAN_STD_RXI}.

The $\hat\sigma_{\text{X},i}^2$ [interference to signal ratio: Eq~\eqref{EQ:RVNORM}]
 values in~\figref{FIG:DECSIG_EXAMPLE}(c) are computed by~\eqref{EQ:ASYNC_MEAN_STD_RXI} and normalized by $4\Delta f^2$ based on statistical procedures. It is found that reducing the bandwidth by half doubles $\hat\sigma_{\text{X},i}^2$ ($\hat\sigma_{\text{X},i}^2=0.0025$ for $10$~GHz and $\hat\sigma_{\text{X},i}^2=0.005$ for $5$~GHz), and hence halves the SIR according to~\eqref{EQ:SIR}, i.e. SIR is linearly proportional to $\Delta f$, as predicted by the last-but-one equality in~\eqref{EQ:SIR2}. Similarly, we have numerically verified that fixing $\Delta f$ and varying $\Delta\tau$ yields SIR linearly proportional to $\Delta\tau$, confirming that SIR is also linearly proportional to $\Delta\tau$, also as predicted by the last-but-one equality in~\eqref{EQ:SIR2}. Thus, SIR is, globally, linearly proportional to DSBP ($\Delta f\Delta\tau$), as indicated by the last equality in Eq.~\eqref{EQ:SIR2}. %Moreover, the values of $\hat\sigma_{\text{X},i}^2$ indicated in the figure are found to be equal to those computed by~\eqref{EQ:MAIPOW_GENERAL1} with $\alpha_i^2=1$ [$\alpha_{ik}^2\equiv1\,(\forall\,i,k)$], $N=2$, $\Delta\tau=10$~ns and $\Delta f =10$ or $5$~GHz, and normalized by $4\Delta f^2$. %Therefore, SIR$_i'$ [Eq.~\eqref{EQ:SIR2}] and $P_{\text{X},i}$ [Eq.~\eqref{EQ:MAIPOW_GENERAL1}] are equivalent to the SIR$_i$ [Eq.~\eqref{EQ:SIR}] and $\sigma_{\text{X},i}^2$ [Eq.~\eqref{EQ:ASYNC_MEAN_STD_RXI}], respectively, for the identical channel energy case ($\alpha_{ik}\equiv1$)

Figure~\ref{FIG:DECSIG_EXAMPLE}(b) plots the normalized DCMA waveforms, $\hat{\tilde{s}}_i(t)=\tilde{s}_i(t)/(2\Delta f)$ and $\hat x_i(t)=x_i(t)/(2\Delta f)$ for $\Delta f=10$~GHz and $\Delta f=5$~GHz to investigate how SIR is affected by the change of bandwidth. It is seen that the normalized MAIs, $\hat x_i(t)$ for $\Delta f=5$~GHz case are larger than their $10$~GHz counterparts, .

\begin{figure}[h!t]
   \centering
   \includegraphics[width=1\columnwidth]{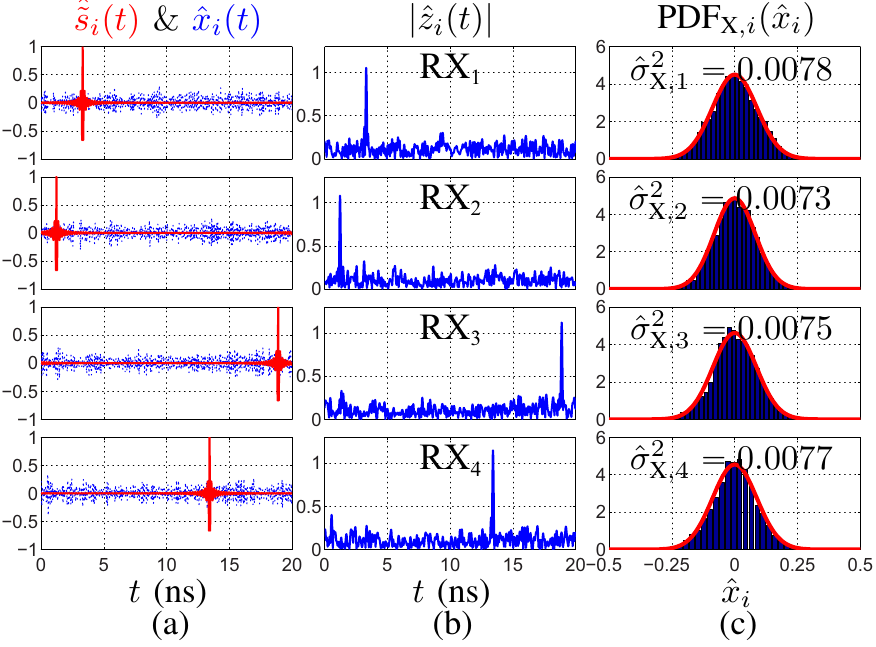}
%   \psfragfig[width=1\linewidth, trim={0in -0.150in  0in -0.2in}]{summai}{
%   \psfrag{Z}[cb][cb][1]{${|\hat{z}_i(t)|} $}
%   \psfrag{S}[cb][cb][1]{$\textcolor{red}{\hat{\tilde{s}}_i(t)}~\&~\textcolor{blue}{\hat{x}_i(t)}$}
%   \psfrag{P}[cb][cb][1]{$\text{PDF}_{\text{X},i}(\hat x_i)$}
%   \psfrag{T}[ct][ct][1]{$t$~(ns)}
%   \psfrag{X}[ct][ct][1]{$\hat x_i$}
%   \psfrag{A}[lt][rt][1]{$\hat\sigma_{\text{X},1}^2=0.0078$}
%   \psfrag{B}[lt][rt][1]{$\hat\sigma_{\text{X},2}^2=0.0073$}
%   \psfrag{C}[lt][rt][1]{$\hat\sigma_{\text{X},3}^2=0.0075$}
%   \psfrag{D}[lt][rt][1]{$\hat\sigma_{\text{X},4}^2=0.0077$}
%   \psfrag{M}[cm][cb][1]{(a)}
%   \psfrag{N}[cm][cb][1]{(b)}
%   \psfrag{O}[cm][cb][1]{(c)}
% }
   \caption{$4\times 4$ DCMA system with $\mathbf{C}=[3,-3,19,-19]$, $\Delta\tau = 10$~ns, $\Delta f = 10$~GHz, $\sigma_\text{N} = 0$ and $\alpha_{ik}^2\equiv1\,(\forall\,i,k)$. (a)~Desired signal, $\hat{\tilde{s}}_i(t)$ (red solid) and MAI, $\hat{x}_i(t)$ (blue dashed) [Eq.~\eqref{EQ:ASYNC_DECSIGS} normalized by $2\Delta f$]. (b)~Decoded signal envelope, $|\hat z_i(t)|$ [Eq.~\eqref{EQ:ASYNC_DECSIG}]. (c)~Actual MAI distribution (bars) and approximated normal distribution [Eq.~\eqref{EQ:ASYNC_GPDF} normalized by $2\Delta f$] with $\hat\sigma_{\text{X},i}^2$ values [interference to signal ratio: Eq.~\eqref{EQ:RVNORM}].}
   \label{FIG:SUMMAI}
\end{figure}

Finally, ~\figref{FIG:SUMMAI} plots the waveforms and MAIs with $\mathbf{C}=[3,-3,19,-19]$, $\Delta\tau=10$~ns and $\Delta f =10$~GHz in a $4 \times 4$ DCMA system. MAI still follows the normal distribution. The corresponding $\hat\sigma_{\text{X},i}^2$ values, computed by~\eqref{EQ:ASYNC_MEAN_STD_RXI}, normalized by $4\Delta f^2$ [Eq.~\eqref{EQ:RVNORM}], and indicated in the figure, agree with those computed by~\eqref{EQ:MAIPOW_GENERAL1} with $\alpha_i^2=1$ [$\alpha_{ik}^2\equiv1\,(\forall\,i,k)$], $N=4$, $\Delta\tau=10$~ns and $\Delta f =10$~GHz and normalized by $4\Delta f^2$, which yields $P_{\text{X},i}^2/(4\Delta f^2) = 0.0075$. %Moreover, it is found, and not graphically shown, that the mixture of $m_i=\pm1$ and the other odd-order codes, i.e. $\mathbf{C}\supset\{1,-1\}$, still yields the total MAI that follows normal distribution.

\subsection{BEP Analysis}

We may now add the AWGN onto MAI, corresponding to the Signal-to-Noise Ratio (SNR)
\begin{equation}\label{EQ:SNR}
  \text{SNR} = \dfrac{4\Delta f^2}{\sigma_\text{N}^2} = \dfrac{1}{\hat\sigma_\text{N}^2},
\end{equation}
where $\sigma_\text{N}^2$ [Eq.~\eqref{EQ:DCMCONCEPT_DECSIGWFM_AWGN}] is the average noise power and $\hat\sigma_\text{N}^2=\sigma_\text{N}^2/(4\Delta f^2)$ is the normalized noise power, correspond the random variable $\hat n$. Assuming, for simplicity, that SNR [Eq.~\eqref{EQ:SNR}] is identical for all receivers (RX$_i$), and that MAI and AWGN are statistically independent, the total interference $\hat u$ (normalized), which is the sum of the two random variables $\hat x_i$ and $\hat n$, i.e. $\hat u = \hat x_i+\hat n$, also follows the normal distribution, i.e.
\begin{equation}\label{EQ:ASYNC_IN_PDF}
  \text{PDF}_i(\hat u)=\dfrac{1}{\sqrt{2\pi(\hat\sigma_{\text{X},i}^2+\hat\sigma_\text{N}^2)}}\exp\left[-\dfrac{\hat u^2}{2(\hat\sigma_{\text{X},i}^2+\hat\sigma_\text{N}^2)}\right],
\end{equation}
with total zero mean and total variance $\hat\sigma^2 = \hat\sigma_\text{X}^2+\hat\sigma_\text{N}^2$. The variance $\hat\sigma^2$ is equivalent to the normalized interference power. Therefore, the Signal to Interference and Noise Ratio (SINR) is
\begin{equation}\label{EQ:ASYNC_SINR}
  \text{SINR}_i = \dfrac{1}{\hat\sigma_{\text{X},i}^2+\hat\sigma_\text{N}^2}=\dfrac{1}{\frac{1}{\text{SIR}_i}+\frac{1}{\text{SNR}}}.
\end{equation}

Assume the worst case case scenario $d_{k,\ell}\equiv1\,(\forall\,k)$ in~\eqref{EQ:ASYNC_DECSIG_MAI}, i.e. all undersired transmitters are sending only 1's, leading to maximal MAI, and assume that the desired signal amplitude is normalized and detected with a threshold of $0.5$. When a bit $d_{i,\ell}=1$ is sent [desired transmitter, Eq.~\eqref{EQ:ASYNC_DECSIG_DESIRED}], $1+\hat x+\hat n<0.5$ ($\hat u<-0.5$) corresponds to an error, whereas when a bit $d_{i,\ell}=0$ is sent (desired transmitter), $0+\hat x+\hat n>0.5$ ($\hat u>0.5$) corresponds to an error. The BEP is thus the following integral of ~\eqref{EQ:ASYNC_IN_PDF} over $\hat u\in(-\infty, -0.5)$ and $\hat u\in(0.5, \infty)$~\cite{BK:1998_Lathi}:
\begin{equation}\label{EQ:ASYNC_BEP}
\begin{split}
   \text{BEP}_i = & \text{PROB}(\hat u<-0.5|d_{i,\ell}=1)+\\
   &\text{PROB}(\hat u>0.5|d_{i,\ell}=0)\\
  =  &0.5\int_{-\infty}^{-0.5}\text{PDF}_i(\hat u)\,d\hat u+\\
  &0.5\int_{0.5}^{-\infty}\text{PDF}_i(\hat u)\,d\hat u,
\end{split}
\end{equation}
where equal transmission probability has been assumed for bits 1 and 0. Inserting \eqref{EQ:ASYNC_IN_PDF} into~\eqref{EQ:ASYNC_BEP} yields after some algebraic manipulations
\begin{subequations}
\begin{equation}\label{EQ:ASYNC_BEP1}
  \text{BEP}_i = \text{Q}\left(\dfrac{\sqrt{\text{SINR}_i}}{2}\right)
\end{equation}
with
\begin{equation}\label{EQ:QFUNC}
  \text{Q}(x) = \dfrac{1}{\sqrt{2\pi}}\int_{x}^{\infty}\exp\left(-\dfrac{x'^2}{2}\right)\,dx'.
\end{equation}
\end{subequations}
Particularly, when $\sigma_\text{N}=0$ (noiseless channel),
\begin{equation}\label{EQ:ASYNC_BEP_N0}
  \text{BEP}_i |_{\sigma_\text{N}=0}=  \text{Q}\left(\dfrac{\sqrt{\text{SIR}_i}}{2}\right),
\end{equation}
with SIR$_i$ given by~\eqref{EQ:SIR} or~\eqref{EQ:SIR2}. The overall BEP is the average of $\text{BEP}_i$ over all receivers, i.e.
\begin{equation}\label{EQ:AVGPBE}
  \text{BEP} =\dfrac{1}{N}\sum_{i=1}^{N}\text{BEP}_i.
\end{equation}

\begin{figure}[h!t]
   \centering
   \includegraphics[width=1\columnwidth]{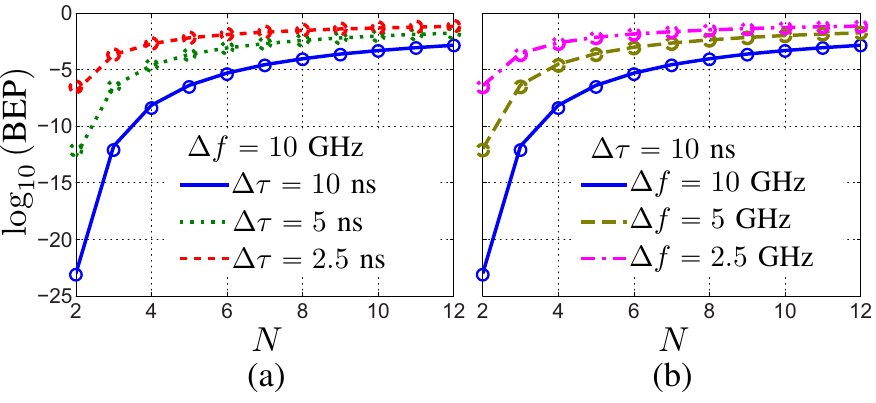}
%   \psfragfig[width=1\linewidth, trim={-0.02in -0.1in  0in 0in}]{bep1}{
%   \psfrag{H}[l][l][0.85]{$\Delta\tau = 10$~ns}
%   \psfrag{E}[l][l][0.85]{$\Delta f=10$~GHz}
%   \psfrag{F}[l][l][0.85]{$\Delta f=5$~GHz}
%   \psfrag{G}[l][l][0.85]{$\Delta f=2.5$~GHz}
%   \psfrag{D}[l][l][0.85]{$\Delta f = 10$~GHz}
%   \psfrag{A}[l][l][0.85]{$\Delta\tau=10$~ns}
%   \psfrag{B}[l][l][0.85]{$\Delta\tau=5$~ns}
%   \psfrag{C}[l][l][0.85]{$\Delta\tau=2.5$~ns}
%   \psfrag{P}[c][c][1]{$\log_{10}\left(\text{BEP}\right)$}
%   \psfrag{N}[c][c][1]{$N$}
%   \psfrag{X}[ct][ct][1]{(a)}
%   \psfrag{Z}[ct][ct][1]{(b)}
% }
   \caption{BEP [bit rate $R_\text{b}=1/(2\Delta\tau)$] versus $N$ for an all-odd DCMA system with  coding~\eqref{EQ:ODD_CODING} $\sigma_\text{N}=0$, $\alpha_{ik}^2\equiv1\,(\forall\,i,k)$. These results are computed by~\eqref{EQ:ASYNC_BEP_N0} and~\eqref{EQ:AVGPBE} with SIR argument given by Eq.~\eqref{EQ:SIR} (Sold lines) or Eq.~\eqref{EQ:SIR2} with $\alpha_i^2=1$ (circle marks). (a)~Fixed bandwidth, $\Delta f=10$~GHz and varying group delay swing, $\Delta\tau$. (b)~Fixed $\Delta\tau=10$~ns and varying $\Delta f$.}
   \label{FIG:BEP_VAR_TBP}
\end{figure}

Figure~\ref{FIG:BEP_VAR_TBP} plots BEP as function of $N$ for a DCMA system with the all-odd coding set
\begin{equation}\label{EQ:ODD_CODING}
\mathbf{C}=\left\{
  \begin{array}{ll}
    [3,-3,\ldots, N+2],  & N~\text{is odd}, \\
    {[}3,-3,\ldots, N+1, -(N+1)],  & N~\text{is even},
  \end{array}
\right.
\end{equation}
in a noiseless channel with unique channel energy [$\alpha_{ik}^2\equiv1\,(\forall\,i,k)$]. This scenario typically resembles that of a downlink communication, where all the channels share a unique transmitting antenna and send signals at the same power level, such that the desired signal and MAI arrive at the receiver with same energy. We see that, in this case, the BEP with argument given by~\eqref{EQ:SIR} and~\eqref{EQ:SIR2} are identical. Changing the group delay swing and changing the bandwidth have identical effect on the BEP. In general, the BEP degrades as $N$ increases, as a result of MAI accumulation and hence degraded SIR according to~\eqref{EQ:SIR2}. Such degradation of SIR can be naturally mitigated by increasing the group delay swing, $\Delta\tau$ [\figref{FIG:BEP_VAR_TBP}(a)]. However, the corresponding bit rate [$R_\text{b}=1/(2\Delta\tau)$] has to be consequently decreased to avoid inter-symbol interference, which degrades the spectral efficiency (bit rate per unit bandwidth) given the assumed fixed bandwidth. Another efficient way to enhance SIR, without affecting the bit rate, is increasing the bandwidth $\Delta f$ [\figref{FIG:BEP_VAR_TBP}(b)]. However, this also degrades the spectral efficiency. In general, the DSBP is the critical figure of merit for DCMA SIR and spectral efficiency, and hence DCMA at large.

\begin{figure}[h!t]
   \centering
   \includegraphics[width=1\columnwidth]{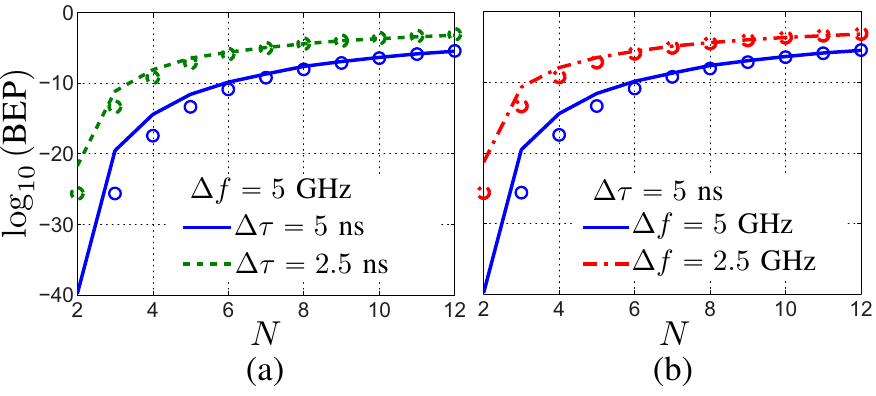}
%   \psfragfig[width=1\linewidth, trim={-0.02in -0.1in  0in 0in}]{bep2}{
%   \psfrag{H}[l][l][0.85]{$\Delta\tau = 5$~ns}
%   \psfrag{E}[l][l][0.85]{$\Delta f=5$~GHz}
%   \psfrag{F}[l][l][0.85]{$\Delta f=2.5$~GHz}
%   \psfrag{D}[l][l][0.85]{$\Delta f = 5$~GHz}
%   \psfrag{A}[l][l][0.85]{$\Delta\tau=5$~ns}
%   \psfrag{B}[l][l][0.85]{$\Delta\tau=2.5$~ns}
%   \psfrag{P}[c][c][1]{$\log_{10}\left(\text{BEP}\right)$}
%   \psfrag{N}[c][c][1]{$N$}
%   \psfrag{X}[ct][ct][1]{(a)}
%   \psfrag{Z}[ct][ct][1]{(b)}
% }
   \caption{BEP [$R_\text{b}=1/(2\Delta\tau)$] versus $N$ for an all-odd DCMA system [Eq.~\eqref{EQ:ODD_CODING}] in a noiseless channel with $\alpha_{ik}^2=\mathcal{U}(0.06, 0.14)$  [Eq.~\eqref{EQ:CHAN_ENGY_DIST} with $\alpha_\text{min}^2=0.06$ and $\alpha_\text{max}^2=0.14$]. These results are computed by~\eqref{EQ:ASYNC_BEP_N0} and~\eqref{EQ:AVGPBE} with SIR argument given by Eq.~\eqref{EQ:SIR} (solid lines) or by Eq.~\eqref{EQ:SIR2} with $\alpha_i^2=(\alpha_\text{min}^2+\alpha_\text{max}^2)/2=0.1$ ($-10$~dB) (circle marks). (a)~Fixed $\Delta f=5$~GHz and varying $\Delta\tau$. (b)~Fixed $\Delta\tau=5$~ns and varying $\Delta f$.}
   \label{FIG:BEP_VAR_TBP2}
\end{figure}

% This condition may be satisfied by ensuring the transmitting signal level at TX$_k$ ($k\neq i$) lower than that TX$_i$, i.e. $g_k<g_i=1\,\forall\,k\neq i$ [Eq.~\eqref{EQ:SYS_SIG}], which may be done by increasing power amplification gain at TX$_i$. However, this measure only may take effect at the RX$_i$, whereas increasing the interference and decreasing the desired signal level at the other receivers RX$_k$ ($k\neq i$), if the antennas are omnidirectional.
In practice, one may further reduce MAI by achieving $\alpha_{ik}^2<\alpha_{ii}^2=1,\,\forall\,k\neq i$ [Eq.~\eqref{EQ:ASYNC_DECSIG_AMP}] with beam forming, which is practically feasible and typically appropriate in a static wireless environment. Figure~\ref{FIG:BEP_VAR_TBP2} plots again BEP but with $\alpha_{ik}^2=\mathcal{U}(0.06, 0.14)$ [Eq.~\eqref{EQ:CHAN_ENGY_DIST}], so as to ensure $\alpha_{ik}^2<\alpha_{ii}^2=1,\,\forall\,k\neq i$. The mean of $\alpha_{ik}^2$ is given by~\eqref{EQ:MAIPOW_GENERAL2}, and approaches $(\alpha_\text{min}^2+\alpha_\text{max}^2)/2=0.1$ ($-10$~dB) as $N$ increases, i.e.
\begin{equation}\label{EQ:MEANENGY}
  \lim_{N\rightarrow\infty} \alpha_{i}^2=\lim_{N\rightarrow\infty}\left(\dfrac{1}{N-1}\sum_{\substack{k=1\\k\neq i}}^{N}\alpha_{ik}^2\right)= \dfrac{\alpha_\text{min}^2+\alpha_\text{max}^2}{2},
\end{equation}
which, according to~\eqref{EQ:SIR2}, ideally (for large $N$) corresponds to SIR enhancement by a factor of $1/\alpha_i^2=10$ ($10$~dB) according to~\eqref{EQ:SIR2}. The BEP results based on Eq.~\eqref{EQ:SIR} (sold lines) essentially agree with those based on~\eqref{EQ:SIR2} with $\alpha_i^2=0.1$ (circle marks). In this scenario, the SIR enhancement via $\alpha_{ik}^2$ decrease allows one to reduce DSBP for a given BEP, which reduces the phaser constraints. %For instance, the BEP at $N=12$ for $\Delta\tau\Delta f = 12.5$ [green dashed line and circles in \figref{FIG:BEP_VAR_TBP2}(a) and red dash-dot line and circles in \figref{FIG:BEP_VAR_TBP2}(b)] is in the order of $10^{-3}$, which is comparable to the BEP at $N=12$ for $\Delta\tau\Delta f = 100$ in~\figref{FIG:BEP_VAR_TBP} (blue lines and circles).

We shall now see that reducing DSBP also leads to increased spectral efficiency at the expense of lower SIR and BEP. The spectral efficiency is defined as the system capacity, $C$, divided by the system bandwidth,$\Delta f$, and is measured in bit per second per Hertz (b/s/Hz). The capacity may be simply expressed as the maximum overall data throughput, $NR_\text{b}=N/T_\text{b}=N/(2\Delta\tau)$, where $R_\text{b}=1/(2\Delta\tau)$ is the maximum allowed bit rate. The spectral efficiency is then
\begin{equation}\label{EQ:SPEC_EFFICIENCY}
  \eta
  =\dfrac{C}{\Delta f}
  =\dfrac{N}{2\Delta\tau \Delta f}
  =\dfrac{N}{2\cdot\text{DSBP}},
\end{equation}
which is inversely proportional to DSBP. Thus, reducing DSBP indeed enhances $\eta$, while degrading SIR and hence the BEP performance, according to~\eqref{EQ:SIR2} and~\eqref{EQ:ASYNC_BEP1}. However, the additional freedom of $\alpha_i^2$, which may be activated by beam forming, allows one to increase $\eta$ without increasing SIR and BEP. Suppose for instance that $N=12$ and that desired spectral efficiency is $\eta = 1$~b/s/Hz, or $\text{DSBP}=\Delta\tau\Delta f=6$, which yields upon insertion into~\eqref{EQ:SIR2} $\text{SIR}=24/(11\alpha_i^2)$. If $\alpha_i^2=1$, $\text{SIR}=2.2$, which is too low for acceptable BEP. Specifying BEP$_\text{max}=0.01$, found to correspond to $\text{SIR}=22$ upon inverting~\eqref{EQ:ASYNC_BEP_N0}, leads to $1/\alpha_i^2\geq 10$ ($10$~dB), which may be achieved using beam forming with proper gains in~\eqref{EQ:ASYNC_DECSIG_AMP}.

\begin{figure}[h!t]
   \centering
   \includegraphics[width=0.7\columnwidth]{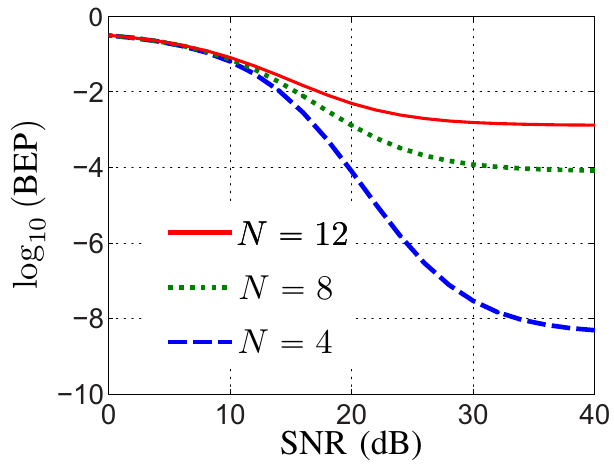}
%   \psfragfig[width=0.7\linewidth, trim={-0.07in 0in  0in 0in}]{bep3}{
%   \psfrag{A}[l][l][1]{$N=12$}
%   \psfrag{B}[l][l][1]{$N=8$}
%   \psfrag{C}[l][l][1]{$N=4$}
%   \psfrag{P}[c][c][1]{$\log_{10}\left(\text{BEP}\right)$}
%   \psfrag{S}[c][c][1]{SNR~(dB)}
% }
   \caption{BEP versus SNR for DCMA with all-odd coding[Eq.~\eqref{EQ:ODD_CODING}], $\Delta f=10$~GHz, $\Delta\tau = 10$~ns in AWGN channel with $\alpha_{ik}^2\equiv1\,(\forall\,i,k)$ [Eq.~\eqref{EQ:CHAN_ENGY_DIST} with $\alpha_\text{min}^2=\alpha_\text{max}^2=1$]. These results are computed by~\eqref{EQ:ASYNC_BEP1} and~\eqref{EQ:AVGPBE} with~\eqref{EQ:ASYNC_SINR} and SIR$_i$ given by~\eqref{EQ:SIR2} with $\alpha_i^2=1$.}
   \label{FIG:BEP_VAR_SNR}
\end{figure}

Figure~\ref{FIG:BEP_VAR_SNR} plots BEP versus SNR for an all-odd DCMA system with $\Delta\tau=10$~ns, $\Delta f = 10$~GHz, and AWGN channels with identical energy. As SNR increases, BEP asymptotically approaches the corresponding noiseless BEP values given in \figref{FIG:BEP_VAR_TBP}, as expected. As SNR decreases below $15$~dB, the three BEP curves converge to the same value, which indicates that in the SNR range $\text{SNR}<15$~dB the total interference is dominated by AWGN. Figure~\ref{FIG:BEP_VAR_SNR} conveniently indicates the SNR requirement for a specified BEP level and DCMA dimension. For instance, for BEP of $10^{-3}$, one finds that SNR should be larger than $16$~dB and $20$~dB for the $N=4$ and $N=8$ cases, respectively, whereas BEP for $N=12$ is always larger than the specified value ($10^{-3}$) even as SNR approaches infinity.

\section{Conclusion}\label{SEC:CONCL}

We theoretically modeled, experimentally demonstrated and statistically characterized Dispersion Code Multiple Access (DCMA) and hence showed the applicability of this purely analog and real-time multiple access scheme to high-speed wireless communications.  We have shown that the critical figure of merit is the Delay Swing Bandwidth Product (DSBP), and that SIR and spectral efficiency are linearly and inversely proportional to the DSBP, respectively. One may consequently use directive antennas to satisfy simultaneously stringent SIR and spectral efficiency specifications. %The self-adaptivity of DCMA system to dynamic communication scenarios may require reconfigurability in the phasers~\cite{JOUR:2017_TMTT_LZOU} or signal routing technologies, such as time-reversal based routing~\cite{CONF:2015_ISAP_LZOU}.

\bibliographystyle{IEEETran}
\bibliography{DCMA_REF}
\end{document}